\documentclass[12pt]{article}
\usepackage[english]{babel}
\usepackage{graphicx,caption,subcaption}
\usepackage{amssymb,amsfonts,amsmath}
\usepackage{indentfirst}
\usepackage{threeparttable}
\usepackage{tablefootnote}
\usepackage{multirow}
\usepackage{ulem}
\usepackage{color}

\usepackage{geometry}
\begin{document}

\title{A mesoscopic model for the effective electrical conductivity of  composite polymeric electrolytes}

\author{M.~Ya.~Sushko, A.~K.~Semenov}

\date{
{\small Department of Theoretical Physics and Astronomy,\\
Mechnikov National University, 2 Dvoryanska St., Odesa 65026,
Ukraine}\\[2ex]
\today }

\maketitle

\begin{abstract}
The effective quasistatic conductivity of composite polymeric
electrolytes is studied in terms of a hard-core--penetrable-layer
model. Used to incorporate the interface phenomena (such as
amorphization of the polymer matrix around filler particles,
stiffening effect by those on the amorphous phase, irregularities
of the filler grains' surfaces, etc.), the layers are assumed to
be electrically inhomogeneous, consisting of a finite or infinite
number of concentric uniform shells. The rules of dominance,
imposed on the overlapping regions, are equivalent to the
requirement that the local electric properties in the system be
determined by the distance from the point of interest to the
nearest particle's center. The desired conductivity is calculated
using our original many-particle (compact-group) approach which,
however, avoids an in-depth elaboration of polarization and
correlation processes. The result is expressed through the
electrical and geometrical parameters of the constituents.
Contrasting it with experiment reveals that the theory adequately
describes the effective conductivity as a function of the filler
concentration and temperature for a number of polymeric composites
based on poly(ethylene oxide) or oxymethylene-linked poly(ethylene
oxide) and is more flexible in comparison with other theories.
\end{abstract}

\section{\label{sec:intro}Introduction}

It has been shown experimentally~\cite{Ploch1988, Przl1995,
Wiec1989, Wiec1994, Moh1998, Zal1996, Siekierski2007} that the
effective electrical conductivity $\sigma_{\rm eff}$ of composite
polymeric electrolytes (CPEs) comprising nano- or microsized
filler particles embedded into a polymer matrix demonstrates a
nonmonotonic behavior with the filler volume concentration $c$.
One typically observes a considerable increase in $\sigma_{\rm
eff}$ at small $c$, even if
nonconductive  particles are added, followed up by a decrease in
$\sigma_{\rm eff}$ at high $c$. The maximum value of $\sigma_{\rm
eff}$ can exceed the electrical conductivity of the pure matrix by
up to three orders of magnitude. This property makes CPEs useful
materials for numerous electrochemical applications, such as
batteries, supercapacitors, fuel cells, sensors, displays, etc.
\cite{MacCallum19871989, Scrosati1993, Bruce1995, Quartarone1998,
Song1999, Croce2000, Sequeira2010, Tarascon2015}.

Qualitatively, the indicated behavior of $\sigma_{\rm eff}$ is
explained by the formation of a highly-conductive amorphous
polymer phase around filler particles due to retardation of the
polymer crystallization process by them~\cite{Knauth2008}. As $c$
is low, the volume fraction of the amorphous regions increases
almost additively, leading to an increase in $\sigma_{\rm eff}$.
As $c$ is high, that decreases, and so does $\sigma_{\rm eff}$.
The existence of such an amorphous phase has been confirmed
experimentally by, say, NMR~\cite{PrzlSymp1988},
DSC~\cite{Przl1995,Bulavin2015}, Raman
spectroscopy~\cite{croce1992}, and X-ray
investigations~\cite{Wiec1989,Ploch1989,Przyluski1990}.

Analytical theories of $\sigma_{\rm eff}$ for systems of spherical
particles are represented by various combinations, such as
\cite{Przl1995,Knauth2008, Nan1993, Jiang1995a, Jiang1995b,
Tod03}, of the classical one-particle approaches by
Maxwell-Garnett~~\cite{Maxwell1873,Maxwell1904,Landau1982} and
Bruggeman (``effective
medium'')~\cite{Bruggeman1935a,Landauer1952} applied to systems of
hard core-shell particles. They pursue the idea
\cite{Nakamura1982,Nakamura1984} that each particle and the
adjacent interphase can be viewed as a single ``complex
particle'', with its effective conductivity found through the
effective response of the isolated complex particle to a uniform
field. In this way, the original three-phase system is reduced to
a two-phase one whose effective conductivity is found again  by
the standard one-particle methods
\cite{Maxwell1873,Maxwell1904,Landau1982,Bruggeman1935a,Landauer1952}
or their modifications. Introducing different quasi-two-phase
models of CPEs for the limiting cases of low and high values of
$c$ \cite{Nan1993,Nan1991L,Nan1991}, and sewing the solutions at
the concentration $c^*$ corresponding to  the maximum of
$\sigma_{\rm {eff}}$, the functional form of $\sigma_{\rm eff}$ is
obtained for the entire range of $c$. As a result, the dependence
of $\sigma_{\rm {eff}}$ upon $c$ has a nonphysical cusp at $c=
c^*$, with $c^*$ being a fitting parameter. Finally, assuming the
validity of the empirical Vogel--Tamman--Fulcher (VTF) equation
(\ref{eq:VTF}) for the shell conductivity and postulating a
certain dependence of $T_0$ upon $c$, the expression for
$\sigma_{\rm eff}$  can be obtained which is
claimed~\cite{Wiec1994} to reproduce both the concentration and
temperature dependences of $\sigma_{\rm eff}$.

In our opinion, repeated applications of homogenization procedure
to the same system, oversimplification of it to one-particle
approaches, and presence of a number of fitting parameters testify
that the above theories cannot be considered as fully consistent.
This fact necessitates the development of new approaches to the
problem.

In this paper, we present a theory for  $\sigma_{\rm eff}$ of CPEs
viewed as  macroscopically homogeneous and isotropic 3D
dispersions of hard-core particles encompassed by penetrable
(freely overlapping) layers. Such models  have already been
proposed and  thoroughly studied in random resistor network
simulations
\cite{Siekierski2007,Knauth2008,Siekierski2005,Siekierski2006}.
Our analytical calculations are based upon the compact-group
approach (CGA) \cite{Sushko2007,Sushko2009,Sushko22009,Sushko2017}
and its results \cite{Sushko2013,Sushko2016} for the case of
homogeneous layers. Even in this simplest case, the theory has
proven to be efficient in describing $\sigma_{\rm eff}$ of
percolating dispersions \cite{Sushko2013} and colloidal
suspensions of nanosized particles \cite{Sushko2016}. Here, we
generalize results \cite{Sushko2013,Sushko2016}  to the case of
inhomogeneous (multi-shell and continuous)  penetrable  layers to
show that the functional forms of the theoretical concentration
and temperature dependences obtained for $\sigma_{\rm eff}$ are
sufficient to describe, in a single way, extensive experimental
data~\cite{Przl1995, Wiec1994} for a series of CPEs based on
poly(ethy1ene oxide) (PEO) and oxymethylene-linked PEO (OMPEO):
PEO--NaI--NASICON  [sodium (Na) Super Ionic CONductor $\rm
Na_{3.2}Zr_2P_{0.8}Si_{2.2}O_{12}$]~\cite{Przl1995},
(PEO)$_{10}$--NaI--$\theta$-Al$_2$O$_3$~\cite{Wiec1994},
PEO--LiClO$_4$--PAAM (polyacrylamid)~\cite{Przl1995, Wiec1994},
and OMPEO--PAAM--LiClO$_4$~\cite{Wiec1994}.

Three crucial features of the theory should be emphasized.

1. We use the notion of penetrable layers as a feasible
mathematical way of modelling the mesoscale structure of real
CPEs. Formally embedding hard-core--penetrable-layer particles
into a uniform matrix and imposing a certain order of dominance
for the overlapping constituents (see
subsection~\ref{subsec:application}), we actually require that the
local properties of the resulting physical system be determined by
the distance from the point of interest to the center of the
nearest particle. In the case of homogeneous layers, we associate
the regions occupied by the layers with a homogeneous amorphous
polymer phase. Finally, taken the layers to be inhomogeneous, say,
consisting of a number of concentric shells, we introduce an
inhomogeneous amorphous phase and other mesostructural units in
real CPEs.

2. For a system with overlapping constituents, the concepts of an
individual particle and its conductivity become ambiguous.
Consequently, neither one-particle classical approaches, nor any
combinations of them are applicable. In contrast, the CGA is a
consistent many-particle approach whose results are expected to be
rigorous in the quasistatic limit. The analysis within the CGA
requires no in-depth elaboration of polarization and correlation
processes, but reduces to simple modeling of the complex
dielectric permittivity distribution in the system, calculations
and summation of the moments of its local deviations from the
effective complex permittivity, and obtaining an integral relation
for $\sigma_{\rm eff}$ in  the quasistatic limit. It will be shown
elsewhere that the CGA results for the $\sigma_{\rm eff}$ of
dispersions of hard-core--penetrable-layer particles are in very
good agreement with the available simulation
results~\cite{Siekierski2007,Knauth2008,Siekierski2005,Siekierski2006}.
Other applications of the CGA can be found in
\cite{Tomylko2015,Semenov2018}, where the problems of step-like
electric percolation  in nematic liquid crystals filled with
multiwalled carbon nanotubes \cite{Tomylko2015} and applicability
of differential mixing rules \cite{Semenov2018} are discussed.

3. The theory is limited to finding and testing the functional
relationships between $\sigma_{\rm eff}$ of CPEs and the
electrical and geometrical parameters of their constituents. It
turns out that the agreement of our results for $\sigma_{\rm eff}$
with experimental data ~\cite{Przl1995, Wiec1994} can be reached
only on condition that the penetrable layer is inhomogeneous. This
fact may signify that several physical and/or chemical mechanisms
are responsible for the formation of the properties of CPEs. Some
suggestions on the relevant physical mechanisms and their effects
on the mesoscale structure of real CPEs  are given in the text.
The feasible chemical mechanisms involving the complexation via
alkali metal cations are discussed in \cite{Wiec1994,Knauth2008}.
Yet the analysis of these questions lies far beyond the scope of
this paper.

The paper is arranged as follows. The basic concepts and relations
of the CGA are presented in subsection~\ref{subsec:basics}. The
model for the microstructure of CPEs and the calculation of
$\sigma_{\rm eff}$ for this model within the CGA are discussed in
subsection~\ref{subsec:application}. The fitting procedures for
comparison of our theory with experiment are outlined in
subsection~\ref{subsec:fittingprocedures}. The results of
processing data~\cite{Przl1995, Wiec1994} for $\sigma_{\rm eff}$
as a function of filler particle concentration and that of
temperature with our theory are discussed in
subsections~\ref{subsec:concentrationdependence} and
\ref{subsec:temperaturedependence}, respectively. The main results
of the paper are summarized in section~\ref{sec:concnlusion}.

\section{\label{sec:compactgroups} Theoretical model}

\subsection{\label{subsec:basics} Basic concepts and relations}
The basic ideas behind the CGA were formulated in
\cite{Sushko2007,Sushko2009,Sushko22009} and developed further in
\cite{Sushko2017,Sushko2013,Sushko2016}. A compact group is
defined as a macroscopic region that contains a sufficiently large
number $N\gg 1$ of structural units (say, filler particles) to
reproduce the properties of the entire system, but still has a
linear size that is much smaller than the wavelength $\lambda$ of
probing radiation. With respect to a field with $\lambda \to
\infty$, a compact group can be treated as a point-like
inhomogeneity, with the fluctuations of $N$ inside being
negligibly small. Under these conditions, a  particulate system
can be viewed as a set of such groups, and its local complex
permittivity can be written as $\hat{\varepsilon}_0 +
\delta\hat{\epsilon}(\bf{r})$, where $\hat{\varepsilon}_0$ is the
complex permittivity of the matrix and
$\delta\hat{\epsilon}(\bf{r})$ is the contribution from a compact
group located at the point of interest $\bf{r}$. The problem is to
find the effective complex permittivity
$\hat{\varepsilon}_{\rm{eff}}$ characterizing the effective
electrodynamic response of the system to a long-wavelength probing
field of frequency $\omega \to 0$.

To allow for different ways of electrodynamic homogenization, we
assume that this response is equivalent to that of a system made
up by embedding the constituents (filler particles and matrix) of
the given system into a uniform host with a complex permittivity
$\hat{\varepsilon}_{\rm f}$. The complex permittivity distribution
in this auxiliary system is modeled as
\begin{equation}\label{eq:profile}
\hat{\varepsilon}(\bf{r}) = \hat{\varepsilon}_{\rm f} + \delta
\hat{\varepsilon}(\bf{r})
\end{equation}
where the term $\delta \hat{\varepsilon}(\bf{r})$ is due to a
compact group (now comprising filler particles and regions
occupied by the real matrix) located at point $\bf {r}$. The
explicit expression for $\delta \hat{\varepsilon}(\bf{r})$ is
constructed in terms of the permittivities and characteristic
(indicator) functions of the constituents, with other relevant
parameters (the degree of penetrability, etc.) taken into account.
For probing fields with the time dependence given, by convention,
by a factor $e^{-{\rm i}\omega t}$, the desired complex
permittivity $\hat{\varepsilon}_{\rm{eff}}$ is defined as the
proportionality coefficient in the relation
\begin{equation} \label{eq:definition}
\langle {\bf{J} (\bf{r})}\rangle = -{\rm i} \omega \langle
{\epsilon_0} \hat{\varepsilon} {(\bf{r}) \bf{E} (\bf{r})} \rangle
= -{\rm i} \omega {\epsilon_0} \hat{\varepsilon}_{\rm eff} \langle
\bf{E} (\bf{r}) \rangle
\end{equation}
where ${\bf{E}}(\bf{r})$ and ${\bf{J}}(\bf{r})$ are the local
field and the complex current density, respectively, the angular
brackets stand for the ensemble averaging or averaging by
integration over the volume, ${\rm i}$ is the imaginary unit, and
$\epsilon_0$ is the electric constant.

In the quasistatic limit $\omega \to 0$, the averages in
Eq.~(\ref{eq:definition}) are formed mainly by multiple
reemissions and correlations within compact groups. As a result,
they can be evaluated, without a detailed elaboration of the
processes involved, by
\begin{equation} \label{eq:field} \langle{\rm {\bf  {E}}}\rangle =
{\left[ {1 + {\sum\limits_{s = 1}^{\infty}  {\left( { -
{\frac{{1}}{{3\hat{\varepsilon}_{\rm f} }} }} \right)^{s} \langle
{{\mathop {\left( {\delta \hat{\varepsilon} ({\rm {\bf r}})}
\right)^{s}}}}} }}\rangle \right]}\,{\rm {\bf E}}_{0}
\end{equation}
\begin{equation}\label{eq:induction}
\langle{\rm {\bf {J}}}\rangle = -{\rm i} \omega \epsilon_0
\hat{\varepsilon}_{\rm f}{\left[ { 1 - 2 {\sum\limits_{s =
1}^{\infty}  {\left( { - {\frac{{1}}{{3\hat{\varepsilon}_{\rm f}
}}}} \right)^{s}{\langle {\mathop {\left( {\delta
\hat{\varepsilon} ({\rm {\bf r}})} \right)^{s}}}}\rangle} }}
\right]}\,{\rm {\bf E}}_{0}
\end{equation}
where ${\bf E}_0$ is the probing field amplitude in the host of
permittivity $\hat{\varepsilon}_{\rm f}$.

So, the procedure for finding $\hat{\varepsilon}_{\rm{eff}}$
involves: 1) the choice of $\hat{\varepsilon_{\rm f}}$; 2)
modeling the permittivity distribution~(\ref{eq:profile}); 3)
finding and summing up the moments of $\delta
\hat{\varepsilon}(\bf{r})$ in Eqs.~(\ref{eq:field}) and
(\ref{eq:induction}). Due to the macroscopicity of compact groups,
the final equation for $\hat{\varepsilon}_{\rm{eff}}$ is
independent of their size.

The approximation $\hat{\varepsilon}_{\rm f}=
\hat{\varepsilon}_0$ is known as the Maxwell-Garnett type of
homogenization. In what follows, we, however,  take
$\hat{\varepsilon}_{\rm f} = \hat{\varepsilon}_{\rm eff}$; this
corresponds to the Bruggeman-type homogenization. Besides being
physically reasonable for systems with complex microstructure, it
is the one that is compatible with the CGA, as shown in
\cite{Sushko2017}. It should be emphasized that despite a seeming
similarity of some of their results, the CGA is not identical to
the classical Maxwell-Garnett
\cite{Maxwell1873,Maxwell1904,Landau1982} and Bruggeman
\cite{Bruggeman1935a,Landauer1952} approaches. The latter deal
with the responses of solitary particles to a uniform field,
whereas the former considers that of a macroscopically large group
of particles embedded in the effective medium of permittivity
$\hat{\varepsilon}_{\rm{eff}}$.

Finally, modeling each complex permittivity involved as
$\hat{\varepsilon} = \varepsilon + {\rm i}{\sigma} /{\epsilon_0
\omega}$, where $\varepsilon$ is the  quasistatic  real part of
the permittivity, and passing to the limit $\omega\to 0$ in the
equation for $\hat{\varepsilon}_{\rm{eff}}$, we obtained a closed
equation for the effective conductivity $\sigma_{\rm{eff}}$ of the
system as a function of the volume concentrations and
conductivities of the constituents.

\subsection{\label{subsec:application} Application to CPEs}

\begin{figure*}[tbh]
\centering
\begin{subfigure}[c]{0.35\textwidth}
\includegraphics[width=\textwidth]{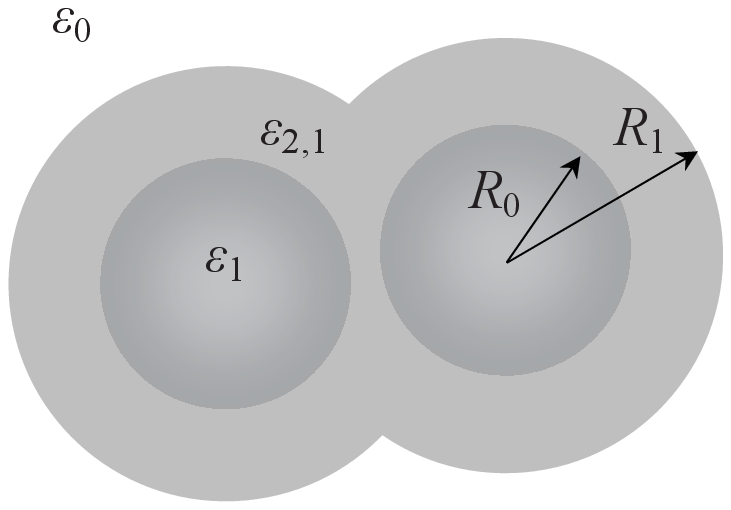}
\caption{} \label{fig:modela}
\end{subfigure}
\begin{subfigure}[c]{0.35\textwidth}
\includegraphics[width=\textwidth]{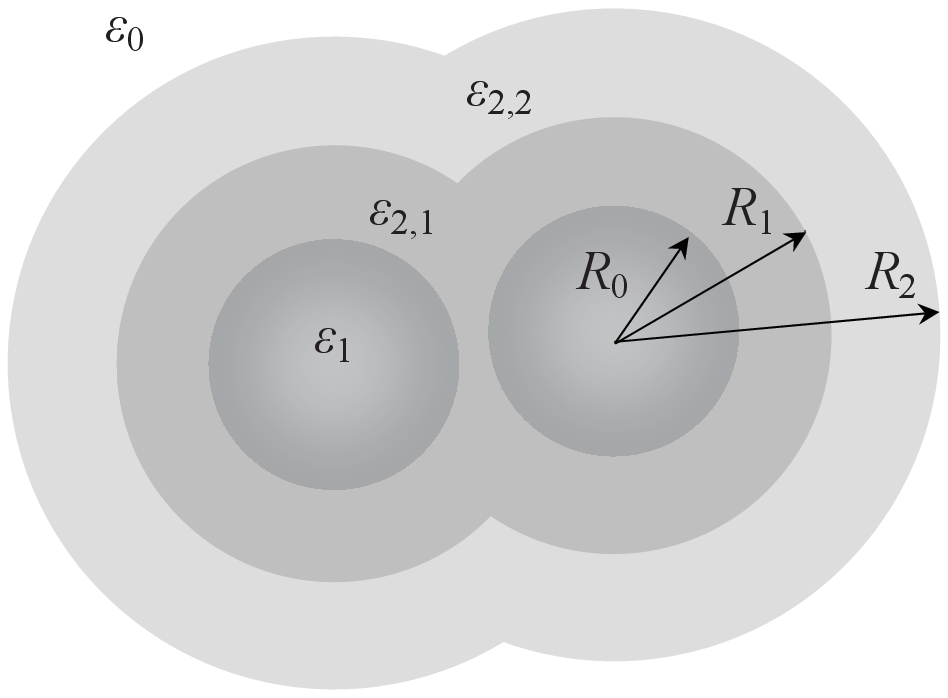}
\caption{} \label{fig:modelb}
\end{subfigure}
\caption{\label{fig:model} A composite electrolyte viewed as a
dispersion of filler particles, with hard cores and penetrable
interphase layers, embedded into a uniform matrix. The cases of
one-shell (homogeneous, a) and two-shell (inhomogeneous, b) layers
are shown. }
\end{figure*}

The model under consideration is partially depicted in
Fig.~\ref{fig:model}. A
CPE is viewed as a dispersion of filler particles
consisting of hard cores, with complex permittivity
$\hat{\varepsilon}_1$ and radius $R_0$, and adjacent interphase
layers; the particles are embedded into a uniform matrix, with
complex permittivity $\hat{\varepsilon}_0$. The interphase layers
are penetrable and in general inhomogeneous. We model each one as
a set of $M$ concentric penetrable shells (an ``$M$-shell''
approximation), with complex permittivities
$\hat{\varepsilon}_{2,j}$ and radii $R_j$, $j=1,\, 2, \dots, M $
($R_0\leq R_1 \leq R_2 \leq \dots \leq R_M$); passing to $M \to
\infty$, we obtain the case of continuous inhomogeneous layers (a
``continuous-layer'' approximation).

To complete the model, a rule defining
the complex permittivity distribution
$\hat{\varepsilon}=\hat{\varepsilon}({\bf {r}})$ in the dispersion
is needed. We assume that the local properties of the dispersion
are formed according to the principle of dominance: if some of its
constituents overlap, the properties of the dominant one are
expressed to the exclusion of those of the others. The order of
dominance is: hard cores $>$ penetrable shells with a smaller $j$
$>$ penetrable shells with a larger  $j$
$>$ the matrix. Consequently, the local value
$\hat{\varepsilon}({\bf {r}})$ is determined only by the distance
$l \equiv \min\limits_{{\bf r}_a} | {\bf r} - {{\bf r}_a} |$ from
the point of interest ${\bf r}$ to the center of the nearest
particle (${{\bf r}_a}$ are the  position vectors of the
particles):
\begin{equation} \hat{\varepsilon}({\bf r})=\begin{cases}
\hat{\varepsilon}_1 & {\text{if} }
\quad \,\,l<R_0\\
\hat{\varepsilon}_{2,j} & {\text{if} }\quad  R_{j-1}<l<R_{j} \,\, (1 \leq j \leq M)\\
\hat{\varepsilon}_0 & {\text{if} }\quad \,\, l>R_M
\end{cases} \label{eq:distr1}
\end{equation}

In view of the accepted rule, the volume concentrations of the
constituents are found as follows. Let $\phi=\phi(c,\delta)$ be
the effective volume concentration (the sum of the hard-core, $c$,
and penetrable-layer, $\phi - c$, volume concentrations) of filler
particles for the case of uniform one-shell layers; here $\delta =
t/R_0$ is the relative thickness of such layers, of thickness
$t=R_1-R_0$, with respect to $R_0$. Given $\phi(c,\delta)$ and
under suggestion (\ref{eq:distr1}), the overall volume
concentration of regions with permittivity
$\hat{\varepsilon}_{2,j}$ is
\begin{equation} \phi_j =\phi(c,\delta_j)-\phi(c,\delta_{j-1})\label{eq:volumej}
\end{equation}
where $\delta_j=(R_j-R_0)/R_0$, $\phi(c,0)=c$.

Thus, provided the fluctuation effects are negligible,  the
dispersion can be considered as an aggregate of non-overlapping
regions with permittivities $\hat{\varepsilon_1}$,
$\hat{\varepsilon}_{2,j}$, $\hat{\varepsilon}_0$ and net volume
concentrations $c$, $\phi_j$, $1 - \phi(c, \delta_M)$,
respectively. Within the CGA, its effective permittivity
$\hat{\varepsilon}_{\rm eff}$ is found as follows. Let $\Pi_0({\bf
r})$, $\Pi_1({\bf r})$ and $\Pi_{2,j}({\bf r})$ be the
characteristic functions of the entire sets of those regions
(occupied by, respectively, the real matrix, hard cores and $j$-th
shells) which form a compact group at point ${\bf r}$ in the
pertinent auxiliary system.  The permittivity distribution in this
system is written in  form
(\ref{eq:profile}) with $\hat{\varepsilon}_{\rm f} =
\hat{\varepsilon}_{\rm eff}$ and, in view of
Eq.~(\ref{eq:distr1}),
\begin{equation}
\delta \hat{\varepsilon}({\bf r })
=(\hat{\varepsilon}_0-\hat{\varepsilon}_{\rm eff}) \Pi_0({\bf r})
+ (\hat{\varepsilon}_1-\hat{\varepsilon}_{\rm
 eff}) \Pi_1({\bf r}) + \sum\limits_{j=1}^M
 (\hat{\varepsilon}_{2,j}-\hat{\varepsilon}_{\rm
 eff}) \Pi_{2,j}({\bf r})
\end{equation}
Due to the orthogonality of $\Pi_0({\bf r})$, $\Pi_1({\bf r})$ and
$\Pi_{2,j}({\bf r})$ to one another, the moments of $\delta
\hat{\varepsilon}({\bf r })$ are calculated readily:
\begin{equation}\label{eq:moments}
    {\langle
{\mathop {\left( {\delta \hat{\varepsilon} ({\rm {\bf r}})}
\right)^{s}}}}\rangle  = \frac{1}{V}\int\limits_V {\mathop {\left(
{\delta \hat{\varepsilon} ({\rm {\bf r}})} \right)^{s}}} d{\bf
r}=(1 - \phi(c,\delta_M) )(\hat{\varepsilon}_0-\hat{\varepsilon}_{\rm
eff})^s + c(\hat{\varepsilon}_1-\hat{\varepsilon}_{\rm eff})^s +
\sum\limits_{j=1}^M \phi_j
(\hat{\varepsilon}_{2,j}-\hat{\varepsilon}_{\rm eff})^s
\end{equation}
Then Eqs.~(\ref{eq:definition}), (\ref{eq:field}), and
(\ref{eq:induction}) give the equation for $\hat{\varepsilon}_{\rm
eff}$:
\begin{eqnarray}
(1-\phi (c,\delta_M ))\frac{\hat{\varepsilon} _0 -
\hat{\varepsilon}_{\rm eff}}{2\, \hat{\varepsilon}_{\rm eff} +
\hat{\varepsilon} _0} + c\frac{\hat{\varepsilon} _1 -
\hat{\varepsilon}_{\rm eff}}{2\,\hat{\varepsilon}_{\rm eff} +
\hat{\varepsilon} _1}  + \sum\limits_{j = 1}^M \phi_j
\frac{\hat{\varepsilon} _{2,j} - \hat{\varepsilon}_{\rm
eff}}{2\,\hat{\varepsilon}_{\rm eff} + \hat{\varepsilon} _{2,j}} =
0 \label{eq: effpermit}
\end{eqnarray}
Whence the desired equation for $\sigma_{\rm eff}$ is obtained:
\begin{eqnarray}
(1-\phi (c,\delta_M ))\frac{\sigma_0 - \sigma_{\rm eff}}{2\,
\sigma_{\rm eff} + \sigma_0} + c\frac{\sigma_1 - \sigma_{\rm
eff}}{2\,\sigma_{\rm eff} + \sigma_1}  + \sum\limits_{j = 1}^M
\phi_j \frac{\sigma_{2,j} -\sigma_{\rm eff}}{2\,\sigma_{\rm eff} +
\sigma_{2,j}} = 0 \label{eq: effconduct}
\end{eqnarray}

For the case of penetrable layers with piecewise-continuous
conductivity profile $\sigma_2=\sigma_2(r)$ and finite thickness,
that is, in the limit $M\to\infty$, $\delta_M$ fixed,
Eq.~(\ref{eq: effconduct}) takes the form
\begin{eqnarray}
(1-\phi (c,\delta_M ))\frac{\sigma_0 - \sigma_{\rm eff}}{2\,
\sigma_{\rm eff} + \sigma_0} + c\frac{\sigma_1 - \sigma_{\rm
eff}}{2\,\sigma_{\rm eff} + \sigma_1}  + \int\limits_0^{\delta_M}
du \frac{\partial \phi (c,u)}{\partial u} \frac{\sigma_2(u)
-\sigma_{\rm eff}}{2\,\sigma_{\rm eff} + \sigma_2(u)} = 0
\label{eq: effconductInhomo}
\end{eqnarray}
where the layer's conductivity profile $\sigma_2(r)$ is expressed
as a function $\sigma_2=\sigma_2(u)$ of the variable
$u=(r-R_0)/R_0$, the relative distance from a current point in the
layer to the surface of the core.

In deriving Eqs.~(\ref{eq: effpermit}) and (\ref{eq: effconduct})
in the above way, the fact of spherical symmetry of cores and
layers is actually insignificant; as a consequence, these
equations remain valid for macroscopically homogeneous and
isotropic dispersions of nonspherical particles, with properly
determined $c$ and $\phi_j$. For spherical
hard-core--penetrable-shell, $\phi=\phi(c,\delta)$ can be
estimated using the scaled-particle
approximation~\cite{Rikvold1985, Rikvold1985a} ($\psi=(1+\delta)^{-3}$, $\phi_t
= c/\psi$):
\begin{eqnarray}
\phi(c,\delta) &=& 1 - (1-c)\exp{\left[ -\frac{(1-\psi)\phi_t}{1-c} \right]}\nonumber\\
&&\times\exp{\left[ -\frac{3c\phi_t}{2(1-c)^3} \left( 2 - 3
\psi^{1/3} + \psi
 - c \left( 3\psi^{1/3} - 6\psi^{2/3} +3\psi \right) \right)
\right]} \label{eq:phi_pen}
\end{eqnarray}
This result is in a good agreement with
Monte Carlo simulations \cite{Lee88,Rottereau03}. We use it in
further applications of the theory.

\section{\label{sec:fitting} Processing experimental data}

\subsection{\label{subsec:fittingprocedures} Fitting procedures}

To put the theory to the test, we first check the
applicability of the $M$-shell and continuous-layer approximations
to the description of the concentration dependence of $\sigma_{\rm
eff}$. Then we use the three-shell approximation to describe the
temperature behavior of $\sigma_{\rm eff}$. The procedures
involved consist of the following three steps:

1. Fitting experimental data~\cite{Przl1995, Wiec1994} for
$\sigma_{\rm eff}$ as a function of $c$  at a fixed temperature
with Eq.~(\ref{eq: effconduct}) in order to determine the relative
thicknesses $\delta_j$ and conductivities $x_{2,j} =
\sigma_{2,j}/\sigma_0$ of the shells in the layer. For each
electrolyte under study, the number of the shells is taken not to
exceed three.

In other words, the one-, two- and three-shell approximations are
exploited for $\sigma_2(r)$ at this step. Outside the hard core
($r>R_0$, $u>0$), the corresponding conductivity profiles $x(u) =
\sigma(r)/\sigma_0$ for the local conductivity values $\sigma(r)$
are, respectively,
\begin{equation}\label{eq:oneshellapproximation}
x(u) = x_{2,1} + (1 - x_{2,1}) \theta(u - \delta_1),\quad M=1
\end{equation}
\begin{equation}\label{eq:twoshellapproximation}
x(u) = x_{2,1} + (x_{2,2} - x_{2,1}) \theta(u - \delta_1) +
(1-x_{2,2}) \theta(u - \delta_2), \quad M=2
\end{equation}
\begin{equation}\label{eq:generalx_triple}
x(u) = x_{2,1} + (x_{2,2} - x_{2,1}) \theta(u - \delta_1) +
(x_{2,3} - x_{2,2}) \theta(u - \delta_2) + (1-x_{2,3}) \theta(u -
\delta_3), \quad M=3
\end{equation}
where $\theta(u)$ is the unit step-function.

If the values of $x_{2,j}$ (or some of them) differ considerably,
then $\sigma_2(r)$ consists of several distinct parts. We assume
that these parts account for different mechanisms (discussed in
subsection \ref{subsec:concentrationdependence}) that are
responsible for the formation of $\sigma_{\rm eff}$ and contribute
most significantly to $\sigma_{\rm eff}$ in certain ranges of $c$.
To estimate these ranges, we take into account the following
facts: in the system of the $j$th shells, with inner radius
$R_{j-1}$ and outer radius $R_j$, percolation paths start to form
at the threshold concentration $c= c_{{\rm c},j}$ given by the
condition $\phi(c_{{\rm c},j},\delta_j) =1/3$ (see
\cite{Sushko2013}); the greatest value of these shells'
contribution to $\sigma_{\rm eff}$ occurs at $c=c_{{\rm m},j}$
where their volume concentration $\phi_j$ has a maximum; the
parameter $x_{2,j}$ governs whether this contribution increases
($x_{2,j}>x_{\rm eff}|_{c_{{\rm c},j}}$) or decreases
($x_{2,j}<x_{\rm eff}|_{c_{{\rm c},j}}$) with $c$ in the interval
$(c_{{\rm c},j},c_{{\rm m},j})$.

Consequently, the behavior of $\sigma_{\rm eff}$ in the interval
$(c_{{\rm c},M}, c_{{\rm m},M})$ is governed by the outermost
shells ($j=M$), with the largest inner and outer radii;
$\sigma_{\rm eff}$ increases if $x_{2,M}> 1$ and decreases if
$x_{2,M}> 1$. As $c$ is further increased, the $(M-1)$th shells,
with smaller inner $R_{M-2}$ and outer $R_{M-1}$ radii, start to
contribute; the role of the $M$th shells starts to diminish; and
the behavior of $\sigma_{\rm eff}$ with $c$ becomes governed by
$x_{2,M-1}$. If, for instance, $x_{2,M-1} \gg x_{2,M} > 1$, then
$\sigma_{\rm eff}$ should keep growing with $c$. However, if
$x_{2,M}>1$ and $ x_{2,M-1} \ll x_{2,M}$, then a maximum  of
$\sigma_{\rm eff}$ is expected to appear near $c_{{\rm m},M}$; and
if $x_{2,M} < 1$ and $x_{2,M-1}\gg 1$, then a minimum of
$\sigma_{\rm eff}$ is expected near $c_{{\rm m},M}$. For
sufficiently large and differing $\delta_M$ and $\delta_{M-1}$,
these local extrema may become resolvable. The parameters of their
location and height (depth) are used to obtain preliminary
estimates for $\delta_M$, $\delta_{M-1}$, and $x_{2,M}$; that for
$x_{2,M-1}$ is obtained by varying $x_{2,M-1}$ so as to recover
the behavior of $\sigma_{\rm eff}$ at $c
> c_{{\rm m},M}$.

Similar considerations can be developed for other inner shells,
should they be needed to incorporate a greater number of the
mechanisms involved. They considerably restrict the ranges for
admissible values of the fitting parameters. Furthermore, the
shapes of fitting curves prove to be very sensitive to the fitting
parameters' variations. The result is that for the observed
nonmonotonic dependences of $\sigma_{\rm eff}$ upon $c$, even
by-hand fitting is efficient to obtain the fitting curves with
reasonable percentage deviations, mostly within the interval
$(-25\%, 25\%)$, from experimental data and high values, $ \approx
0.92$ to 0.99, for the coefficient of determination $R^2$. In the
situation where the number of experimental points is limited and
the experimental errors have not been reported, a further
optimization of the fitting procedure for profiles
(\ref{eq:twoshellapproximation}) and (\ref{eq:generalx_triple})
does not seem necessary.

2. Fitting the same data for $\sigma_{\rm eff}$  using
Eq.~(\ref{eq: effconductInhomo}) with $\sigma_2(r)$ generated by
the continuously differentiable sigmoid-type functions
($0<u<\delta_M$)
\begin{equation}\label{eq:GeneratingFunction2}
x_2(u)= X_{21} +
\frac{X_{22}-X_{21}}
{1+\exp\left[-\frac{u-\Delta_1}{\alpha}\right]}
+\frac{1-X_{22}}{1+\exp\left[-\frac{u-\Delta_2}{\alpha}\right]}
\end{equation}
\begin{equation}\label{eq:GeneratingFunction3}
x_2(u)= X_{21} +
\frac{X_{22}-X_{21}}
{1+\exp\left[-\frac{u-\Delta_1}{\alpha}\right]}
+\frac{X_{23}-X_{22}}{1+\exp\left[-\frac{u-\Delta_2}{\alpha}\right]}
+\frac{1-X_{23}}{1+\exp\left[-\frac{u-\Delta_3}{\alpha}\right]}
\end{equation}
In the limiting case $\alpha\to 0$ and $\delta_M \to \infty$,
profiles~(\ref{eq:GeneratingFunction2}) and
(\ref{eq:GeneratingFunction3}) transform into
profiles~(\ref{eq:twoshellapproximation}) and
(\ref{eq:generalx_triple}), respectively, with $\Delta_j =
\delta_j$ and $X_{2,j}=x_{2,j}$ having the previous physical
meanings. For $\alpha \neq 0$, $\Delta_j$ and $X_{2,j}$ should be
viewed only as formal parameters in the generating
functions~(\ref{eq:GeneratingFunction2}) and
(\ref{eq:GeneratingFunction3}).

At this step, the continuous-layer approximation is used in order
to fit the experimental $\sigma_{\rm eff}$ versus $c$ data. In
doing so, the layer's conductivity profiles of types
(\ref{eq:GeneratingFunction2}) and (\ref{eq:GeneratingFunction3})
are varied by increasing $\alpha$ as much as possible and
adjusting $\Delta_j$ and $X_{2,j}$ properly.

3. Applying the three-shell approximation, with corresponding
$\delta_j$ taken to be temperature-independent, to $\sigma_{\rm
eff}$ versus $c$ data~\cite{Wiec1994} for three conductivity
isotherms of blends of amorphous OMPEO with PAAM in order to
obtain the values for $\sigma_{2,j}$ and $\sigma_0$ at three
different temperatures. Next, assuming that the latter
conductivities obey the VTF equation
\begin{equation}\label{eq:VTF}
\sigma = \frac{A}{\sqrt{T}} \exp{\left( -\frac{B}{T-T_0} \right)}
\end{equation}
the parameters $A$, $B$, and $T_0$ for each shell and the matrix
are found by solving the pertinent systems of three equations.
Finally, having $\delta_j$ from the same fitting procedure, and
disregarding the temperature dependence of $\sigma_1$, the
temperature dependence of $\sigma_{\rm eff}$ is recovered with
Eq.~(\ref{eq: effconduct}) and contrasted with experiment.

\subsection{\label{subsec:concentrationdependence} Concentration dependence of $\sigma_{\rm eff}$}

\begin{figure*}[tbh]
\centering
\begin{subfigure}[c]{0.45\textwidth}
\includegraphics[width=\textwidth]{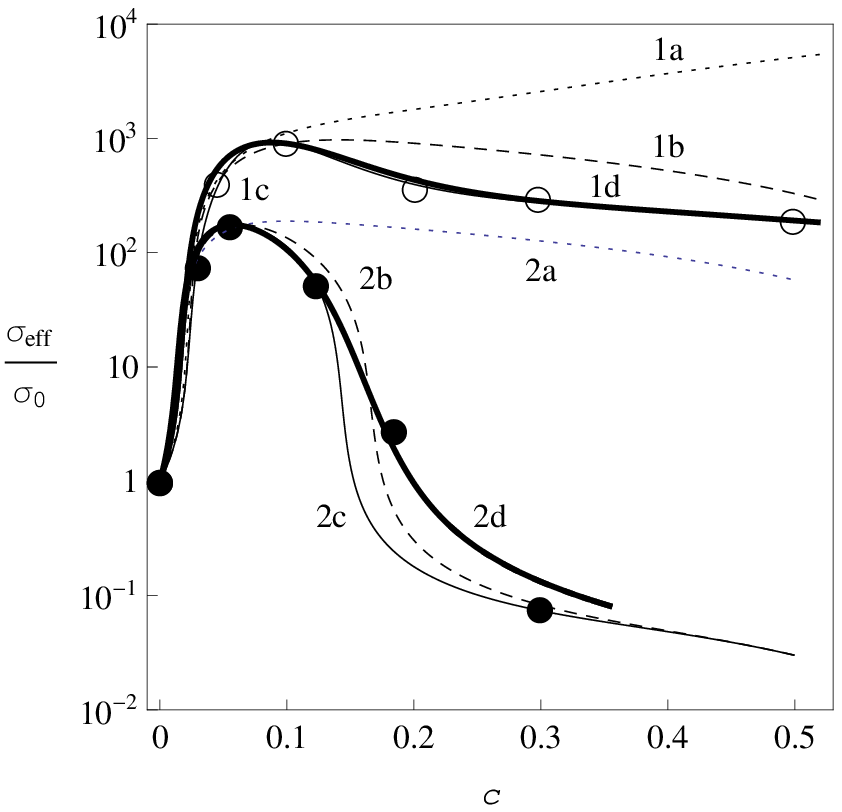}
\caption{} \label{fig:PEO-NaIa}
\end{subfigure}%
~
\begin{subfigure}[c]{0.38\textwidth}
\includegraphics[width=\textwidth]{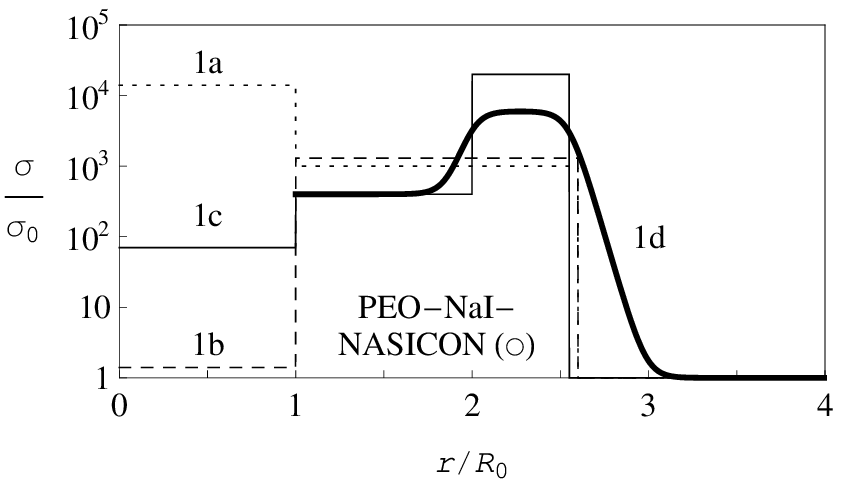}
\includegraphics[width=\textwidth]{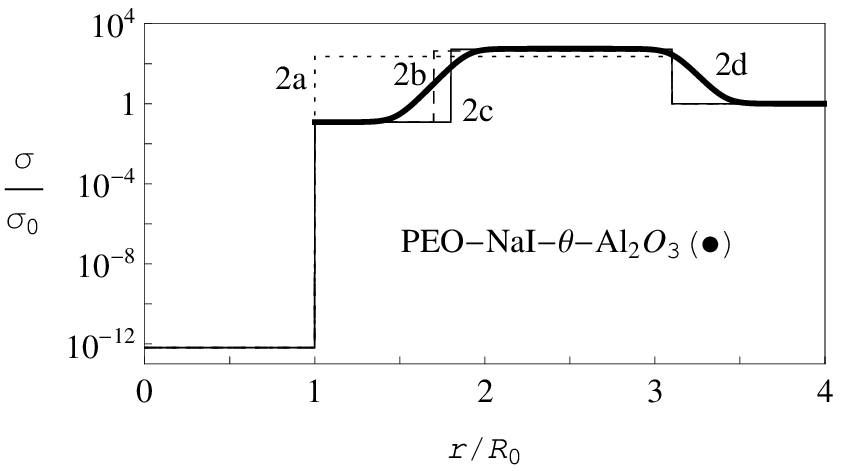}
\caption{} \label{fig:PEO-NaIb}
\end{subfigure}
\caption{\label{fig:PEO-NaI} (a) Experimental data for
$\sigma_{\rm eff}$ as a function of $c$ for PEO--NaI--NASICON
\cite{Przl1995} ($\circ$) and
(PEO)$_{10}$--NaI--$\theta$-Al$_2$O$_3$ \cite{Wiec1994}
($\bullet$) electrolytes, and their fits within the one-shell,
two-shell, and continuous layer approximations. The legends
identify the fits with parameters specified in
Table~\ref{tab:adjustable_params}. (b) The corresponding
one-particle's conductivity profiles used to model the mesoscopic
structure of these electrolytes.}
\end{figure*}

\begin{figure*}[h!tb]
\centering
\begin{subfigure}[c]{0.45\textwidth}
\includegraphics[width=\textwidth]{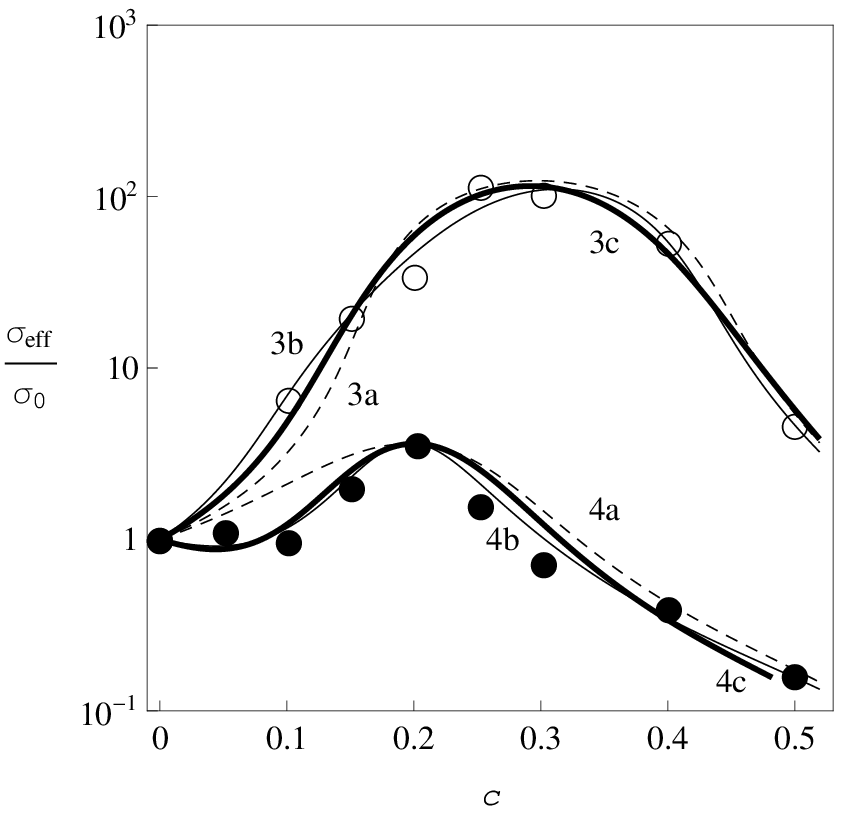}
\caption{} \label{fig:OMPEO-LiClO4a}
\end{subfigure}%
~
\begin{subfigure}[c]{0.38\textwidth}
\includegraphics[width=\textwidth]{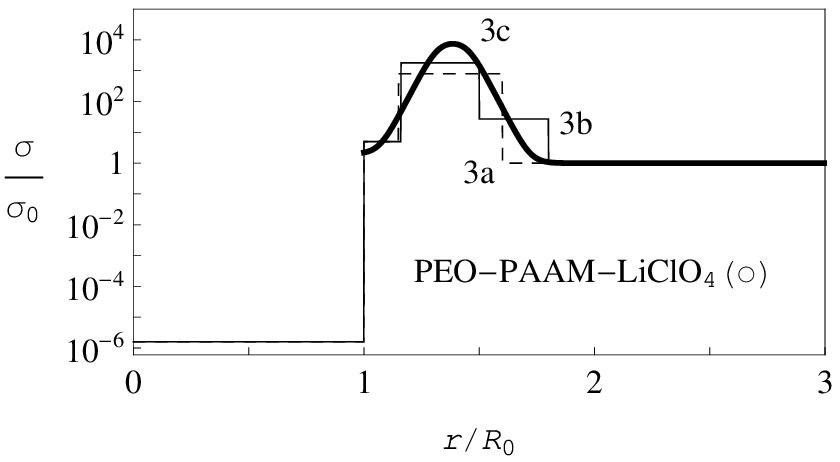}
\includegraphics[width=\textwidth]{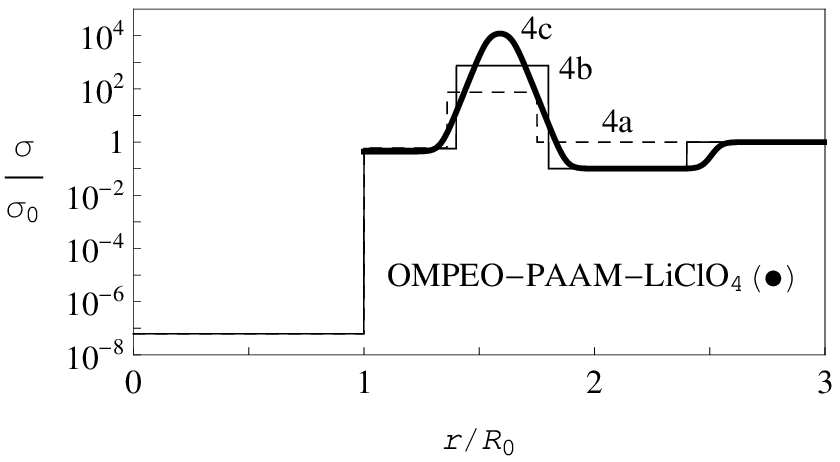}
\caption{} \label{fig:OMPEO-LiClO4b}
\end{subfigure}
\caption{\label{fig:OMPEO-LiClO4} (a) Experimental data for
$\sigma_{\rm eff}$ as a function of $c$ for PEO--PAAM--LiClO$_4$
\cite{Przl1995, Wiec1994} ($\circ$) and OMPEO--PAAM--LiClO$_4$
\cite{Wiec1994} ($\bullet$) electrolytes, and their fits within
the two-shell, three-shell, and continuous layer approximations.
The legends identify the fits with parameters specified in
Table~\ref{tab:adjustable_params}. (b) The corresponding
one-particle's conductivity profiles used to model the mesoscopic
structure of these electrolytes.}
\end{figure*}

\begin{table*}[hbt]
\centering \caption{\label{tab:adjustable_params} Parameters used
to fit $\sigma_{\rm eff}$ vs $c$ data \cite{Przl1995,Wiec1994} for
composite polyether-based electrolytes at $t= 25\,\rm{^oC}$ within
several-shell and continuous-layer approximations, and $R^2$
values for appropriate fits.}
\begin{threeparttable}
\begin{tabular}{lllllllllll}
\hline
\multirow{2}{*}\textrm{ Layer} &L\tnote{a} &  $\sigma_0$, S/cm  & $x_1$& $\delta_1$\tnote{b} & $\delta_2$\tnote{b} & $\delta_3$\tnote{b} & $x_{21}$\tnote{b} & $x_{22}$\tnote{b} & $x_{23}$\tnote{b} &  $R^2$, \% \\
 &  & & & $\Delta_1$\tnote{c}& $\Delta_2$\tnote{c}&$\Delta_3$\tnote{c}&$X_{21}\tnote{c}$&$X_{22}$\tnote{c}&$X_{23} \tnote{c}$ & \\
 \hline
 \multicolumn{11}{c}{PEO--NaI--NASICON}\\
one-shell  &1a   &$9.86\times 10^{-9}$&$1.4\times 10^4$&1.6& -- & -- & 1000& -- & -- & -- \\
one-shell                 &1b                   &                    &1.4             &1.6& -- & -- & 1300& -- & -- & -- \\
two-shell                 &1c                   &                    &70              &1.0&1.55& -- & 400  &  20000 & -- &  99.4 \\
continuous, &1d &                  &70              &1.0&1.55& -- & 400 &  6000 & -- &  95.5 \\
$\alpha =0.05$ &  &  & & & & & &  &  \\
\hline
 \multicolumn{11}{c}{(PEO)$_{10}$--NaI--$\theta$-Al$_2$O$_3$}\\
one-shell &2a&$1.54\times 10^{-8}$&$6.5\times 10^{-13}$&2.1&--&--&230&--&-- & -- \\
two-shell &2b  &                                     &                   &0.7&2.1&--&0.12&435&--& 92.8\\
two-shell &2c  &                                     &                   &0.8&2.1&--&0.12&520&-- & 98.6\\
continuous, &2d&                                  &                   &0.9&2.1&--&0.12&560&--& 95.0\\
$\alpha =0.05$ &  &  & & & & & &  &  &  \\
 \hline
 \multicolumn{11}{c}{PEO--PAAM--LiClO$_4$}\\
two-shell &3a &$6.12\times 10^{-7}$&$1.6\times 10^{-6}$&0.15&0.60& -- & 5.0           &800 &--& 88.7 \\
three-shell  &3b                 &                    &                  &0.16&0.50&0.80 & 5.0 &1800 &27 &  92.3 \\
continuous,  &3c  &                    &                   &0.32&0.45&0.48 & 2.0     &9400&27 & 92.9 \\
$\alpha =0.03$ &  &  & & & & & &  & & \\
\hline
 \multicolumn{11}{c}{OMPEO--PAAM--LiClO$_4$, after annealing}\\
two-shell   &4a           &$1.61\times 10^{-5}$&$6.2\times 10^{-8}$&0.36&0.75& -- &0.60& 75 &--& 46.3\\
three-shell   & 4b &                 &                    &0.40&0.80&1.40&0.57& 750&0.10 &  93.8\\
continuous, &4c               &      &                    &0.54&0.64&1.53&0.44&14200&0.10 &  81.7\\
$\alpha =0.02$ &  &  & & & & & &  & & \\
\hline
\end{tabular}
\begin{tablenotes}
\item[a] Legends used to identify  the corresponding fitting
curves.
\item[b] Parameters for several-shell approximations.
\item[c] Parameters for continuous-layer approximations.
\end{tablenotes}
\end{threeparttable}
\end{table*}

\begin{figure*}[tbh]
\centering
\includegraphics[width=0.85\textwidth]{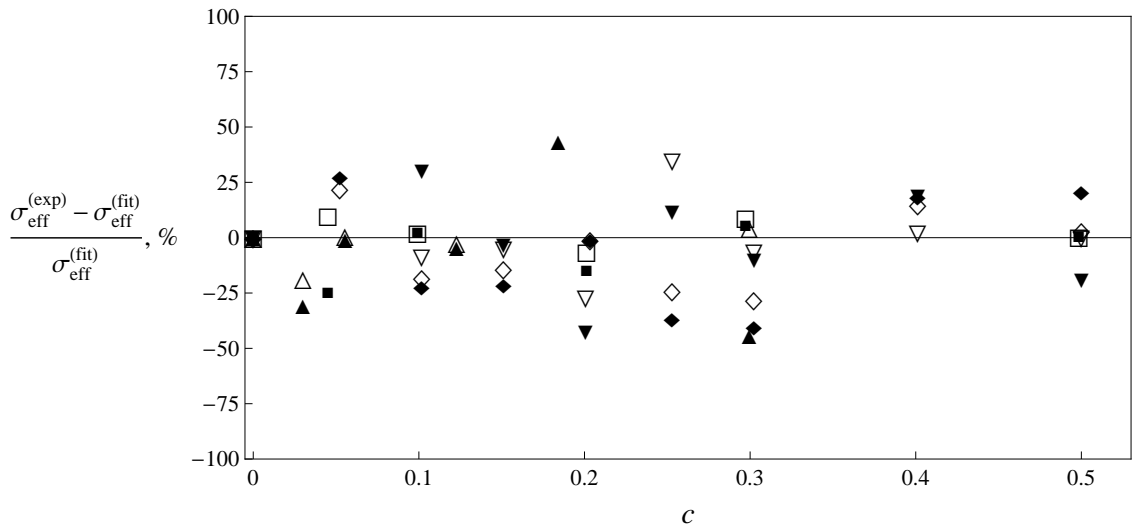}
\caption{\label{fig:RAE-conc}  Percentage deviations of
experimental $\sigma_{\rm eff}$ vs $c$ data for PEO--NaI--NASICON
\cite{Przl1995}, (PEO)$_{10}$--NaI--$\theta$-Al$_2$O$_3$
\cite{Wiec1994}, PEO--PAAM--LiClO$_4$ \cite{Przl1995, Wiec1994},
and OMPEO--PAAM--LiClO$_4$ \cite{Wiec1994} electrolytes from
fitting curves 1c ($\square$), 2c ($\triangle$), 3b
($\triangledown$), and 4b ($\lozenge$), respectively (see
Figs.~\ref{fig:PEO-NaI} and~\ref{fig:OMPEO-LiClO4}). The filled
markers: the same for fitting curves 1d, 2d, 3c, and 4c,
respectively. The percentage deviation, $ \approx 1040\, \%$, for
the reported experimental point $c \approx 0.18$, $x_{\rm eff}
\approx 2.8$ for (PEO)$_{10}$--NaI--$\theta$-Al$_2$O$_3$ (the
fifth $\bullet$ in Fig.~\ref{fig:PEO-NaI}) from the 2c curve is
not shown. The $R^2$ values for the above fits are summarized in
Table~\ref{tab:adjustable_params}. }
\end{figure*}

\begin{figure}[htb]
\centering
\includegraphics[width=0.45\textwidth]{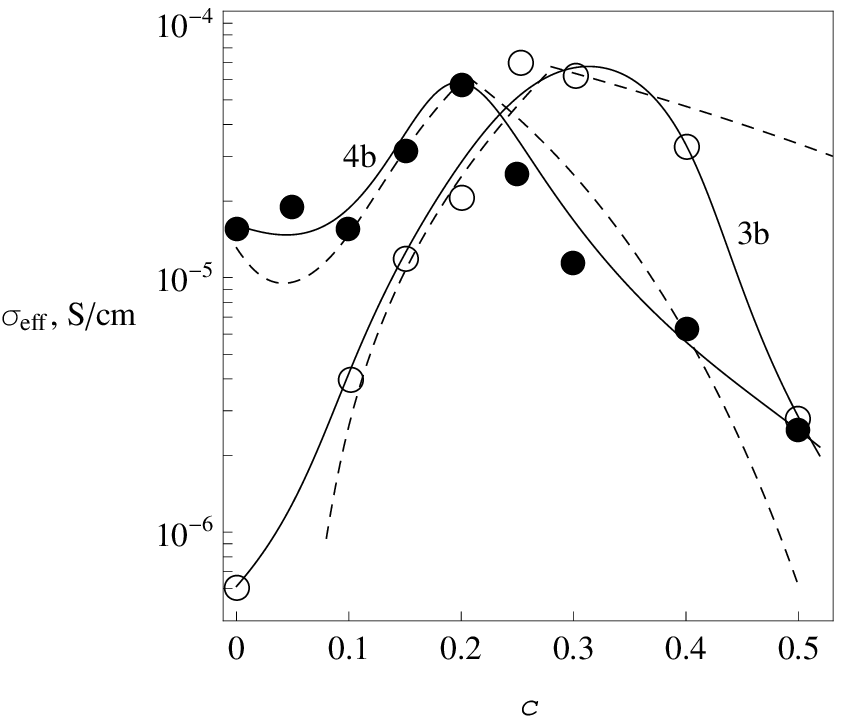}
\caption{\label{fig:comparison} Comparison of the three-shell
approximation (solid lines 3b and 4b, see
Table~\ref{tab:adjustable_params}) with effective medium
theory~\cite{Wiec1994} (dashed lines, see Table~7 and Fig.~10 in
\cite{Wiec1994}), both applied to experimental
data~\cite{Wiec1994} for PEO--PAAM--LiClO$_4$ ($\circ$) and
OMPEO--PAAM--LiClO$_4$ (after annealing) ($\bullet$) at $25 {\rm
^oC}$ (concentration of LiClO$_4$ is equal to 10~mol~\% with
respect to ether oxygen concentration).}
\end{figure}

Experimental data~\cite{Przl1995, Wiec1994} for several types of
PEO- and OMPEO-based CPEs reveal a nonmonotonic behavior of
$\sigma_{\rm eff}$ with $c$, with a maximum of $\sigma_{\rm eff}$
occurring for $c$ in between 0.05 and 0.1 for PEO--NaI--NASICON
and (PEO)$_{10}$--NaI--$\theta$-Al$_2$O$_3$ (see
Fig.~\ref{fig:PEO-NaIa}), that in between 0.2 and 0.3 for
PEO--PAAM--LiClO$_4$ and OMPEO--PAAM--LiClO$_4$
(Fig.~\ref{fig:OMPEO-LiClO4a}),  and possibly a minimum of
$\sigma_{\rm eff}$ at $c$ close to 0.1 for OMPEO--PAAM--LiClO$_4$.
Our fitting results (see Figs.~\ref{fig:PEO-NaIa},
\ref{fig:OMPEO-LiClO4a} and Table~\ref{tab:adjustable_params}) for
different types of $\sigma_2(r)$ (Figs.~\ref{fig:PEO-NaIb} and
\ref{fig:OMPEO-LiClO4b}, respectively) give a clear hint at the
inhomogeneity of it: a good agreement of our theory with
data~\cite{Przl1995, Wiec1994} (see Fig.~\ref{fig:RAE-conc} for
the percentage deviations of experimental data from our fits and
Table~\ref{tab:adjustable_params} for the corresponding values of
$R^2$) is reached within the two-shell approximation for CPEs with
inorganic conductive (NASICON) or nonconductive ($\rm
\theta$-Al$_2$O$_3$) filler particles, and the three-shell
approximation for blends, complexed with LiClO$_4$, of PEO or
OMPEO with PAAM. The values of $\delta_j$ and $x_{2,j}$ obtained
correlate well with those expected to lead to the previously
predicted scenarios of the behavior of $\sigma_{\rm eff}$ with
$c$.

The use of the continuous-layer approximation, which may seem more
adequate physically, alters the model shapes of the one-particle's
conductivity profiles noticeably; in fact, they become similar to
those suggested in
simulations~\cite{Knauth2008,Siekierski2005,Siekierski2006}.
However, at least for the indicated CPEs, this approximation does
not change much the theoretical estimates for $\sigma_{\rm eff}$
as compared to those given by the three-shell approximation, yet
usually reducing the $R^2$ values for the corresponding fits to
the experiment. This fact gives grounds to apply the three-shell
approximation to the study of the temperature behavior of
$\sigma_{\rm eff}$ (see
subsection~\ref{subsec:temperaturedependence}). In a more
comprehensive sense, it sustains the idea that $\sigma_2(r)$
effectively accounts for the net effect on $\sigma_{\rm eff}$ by
different mechanisms.

Based on  the conductivity values obtained (see
Table~\ref{tab:adjustable_params}),  we can assume that
$\sigma_{\rm eff}$ of the above CPEs  is contributed by several
common mechanisms~\cite{Knauth2008}:

1. The amorphization of the polymer matrix by filler grains, that
is, formation of a highly conductive (due to its large disorder
and enhanced ionic mobility) amorphous  polymer phase near the
polymer-filler interface. This effect is attributed to the
inhibition of the polymer crystallization process near the filler
grains, which act as both nucleation centers for the polymer
matrix and mechanical hindrances for polymer crystallite growth.

2. The stiffening effect of  the filler on the amorphous phase,
that is, the reduction of the flexibility of polymer chain
segments and, consequently, of the ionic mobility in the close
vicinity of the polymer-filler interface. It leads to a decrease
in the local conductivity values ($\sigma_{2,1}$) there in
comparison with those ($\sigma_{2,2}$) at greater distances from
the interface. The innermost shell, of conductivity
$\sigma_{2,1}$, is also supposed to incorporate the effects caused
by irregularities in the shape of the filler grains (such as PAAM
globes in polymer blends).

3. An effective decrease, compared to that of the pure filler, of
the conductivity of highly-conductive filler grains inside CPEs
caused by the formation of the highly-resistive polymer-filler
interface. Our processing results for  $\sigma_1$ in
PEO--NaI--NASICON  electrolytes agree with this expectation.

Of interest is the fact that the layer's conductivity profiles
inferred for the OMPEO-based electrolytes exhibit a peak followed
by a trough, whereas those for the PEO-based electrolytes exhibit
a peak alone (see Figs.~\ref{fig:PEO-NaIb} and
\ref{fig:OMPEO-LiClO4b}). To explain it, we should return to the
concept of penetrable layers and emphasize that their conductivity
profiles are not equivalent to the actual conductivity
distributions around the hard  cores, but represent a convenient
way for modeling the effective microstructure of CPEs. The
electrical properties of the layer's outermost part determine the
behavior of $\sigma_{\rm eff}$ at low $c$ where $\sigma_{\rm eff}$
is significantly contributed by the matrix. If the pure matrix
polymer is relatively high-conductive (such as amorphous OMPEO
compared to semicrystalline PEO), then the addition of a
low-conductive polymer (such as PAAM) to it may detectably
decrease its conductivity (say,  due to the formation of complexes
involving the ${\rm Li^{+}}$ cations and PAAM). A visible trough
may then be needed on  $\sigma_2(r)$ to incorporate this effect
within our approach. As $c$ is increased, the interphase amorphous
regions with higher conductivities come into play.

Finally, two examples of contrasting our results
with those predicted by theory~\cite{Wiec1994} (outlined in
section~\ref{sec:intro}) are given in Fig.~\ref{fig:comparison}.
They clearly demonstrate that our approach exhibits a greater
flexibility in the quantitative description of the
electrical conductivities of CPEs.

\subsection{\label{subsec:temperaturedependence} Temperature dependence of $\sigma_{\rm eff}$}

\begin{figure}[htb]
\centering
\includegraphics[width=0.45\textwidth]{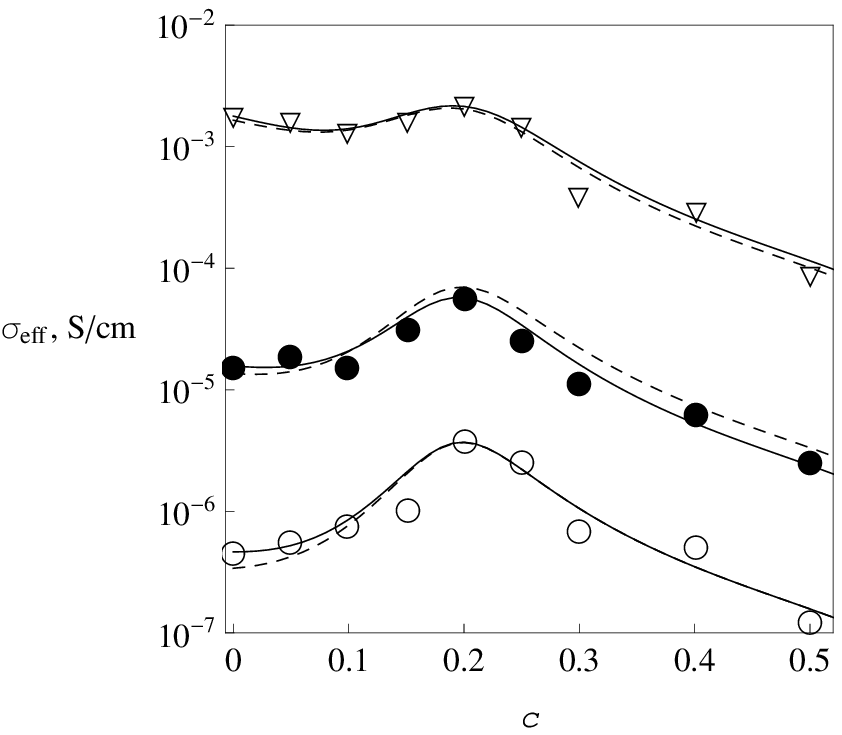}
\caption{\label{fig:OMPEO-LiClO4-Temp} Experimental
data~\cite{Wiec1994} for the $t=0 {\rm ^oC}$ ($\circ$), $25 {\rm
^oC}$ ($\bullet$), and $100 {\rm ^oC}$ ($\nabla$) isotherms of
$\sigma_{\rm eff}$ of OMPEO--PAAM--LiClO$_4$ electrolytes (with 10
mol\% LiClO$_4$, after annealing) versus PAAM concentration.
Dashed lines: VTF-type fits~\cite{Wiec1994} with parameter values
from Table~5 in \cite{Wiec1994}. Solid lines: our fits within the
three-shell approximation, given by Eq.~(\ref{eq: effconduct}) at
$M=3$ [see also Eqs.~(\ref{eq: effconductInhomo}) and
(\ref{eq:generalx_triple})], with parameter values summarized in
Table~\ref{tab:isotherms}. The percentage deviations of these data
from and $R^2$ values for our fits are presented in
Fig.~\ref{fig:OMPEO-LiClO4-Temp-Deviation}.}
\end{figure}

\begin{figure}[htb]
\centering
\includegraphics[width=0.85\textwidth]{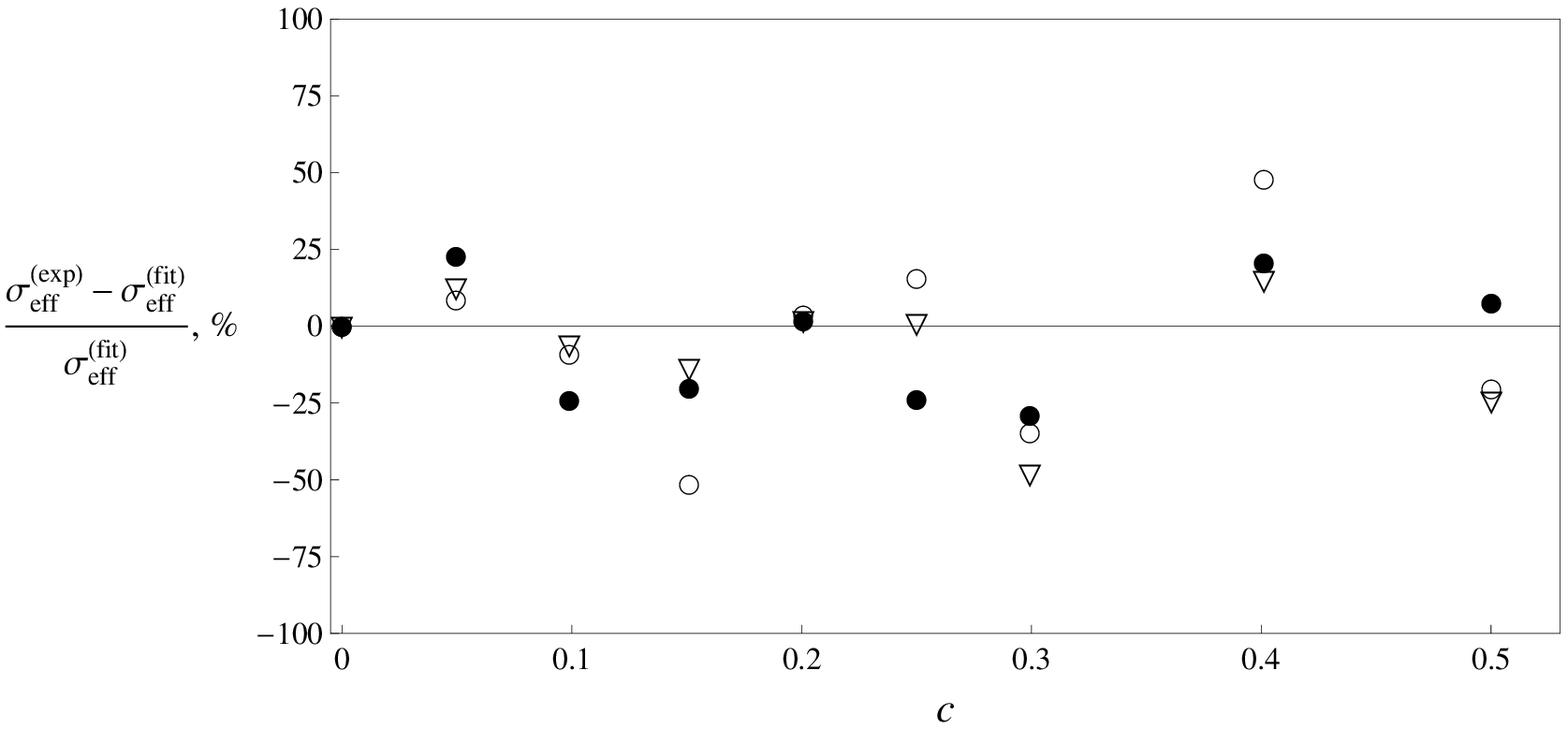}
\caption{\label{fig:OMPEO-LiClO4-Temp-Deviation} Percentage
deviations of experimental $\sigma_{\rm eff}$ vs $c$
data~\cite{Wiec1994} for three $\sigma_{\rm eff}$ isotherms of
OMPEO--PAAM--LiClO$_4$ electrolytes (with 10 mol\% LiClO$_4$,
after annealing) from our fits shown in
Fig.~\ref{fig:OMPEO-LiClO4-Temp}. The markers $\circ$, $\bullet$,
and $\nabla$ refer to the same data as those in
Fig.~\ref{fig:OMPEO-LiClO4-Temp}. The $R^2$ values for these fits
are 87.2, 91.4, and 94.5\,\%, respectively.}
\end{figure}

\begin{figure}[htb]
\centering
\includegraphics[width=0.45\textwidth]{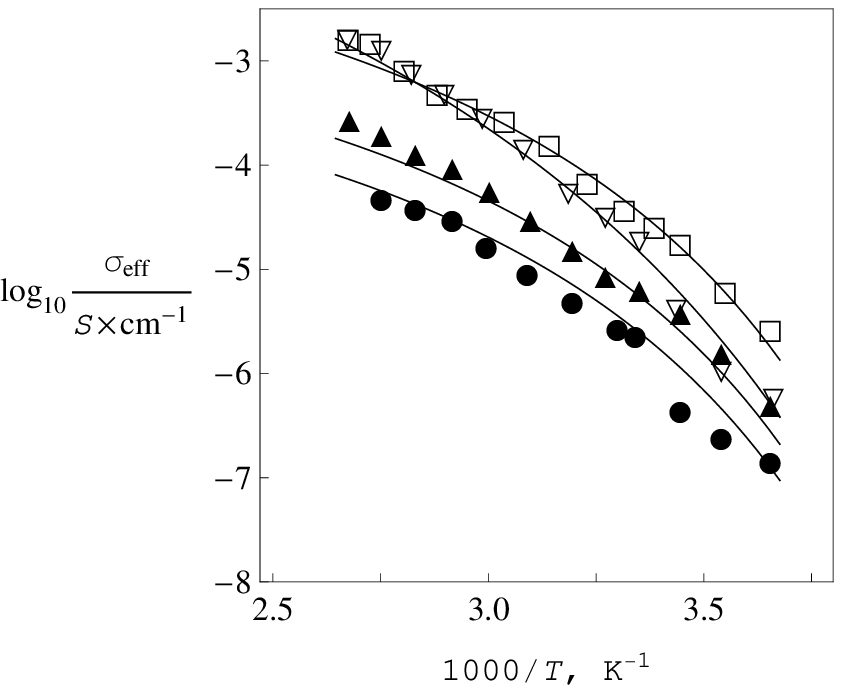}
~
\includegraphics[width=0.45\textwidth]{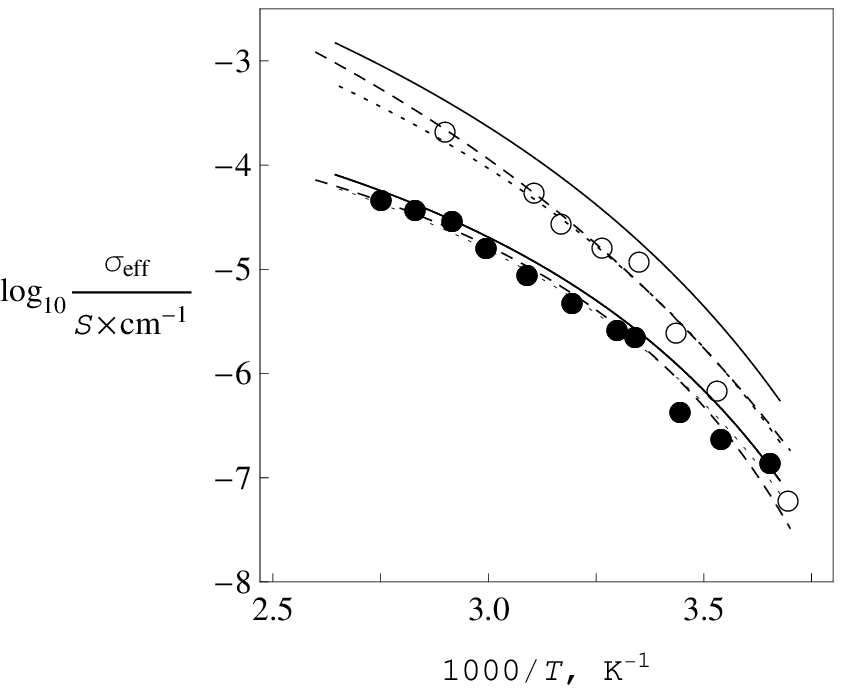}\\
(a)\hfil(b)\hfil \caption{\label{fig:OMPEO-LiClO4-TempDependence}
Experimental data~\cite{Wiec1994} for $\sigma_{\rm eff}$ as a
function of temperature for OMPEO--PAAM--LiClO$_4$ electrolytes
(10 mol\% LiClO$_4$, after annealing) with 5 ($\nabla$), 10
($\circ$), 25 ($\square$), 40 ($\blacktriangle$), and 50
($\bullet$) vol\% PAAM. Dashed lines, (b): VTF-type fits, with
parameter values from Table~5 in \cite{Wiec1994}, proposed by the
authors of \cite{Wiec1994} for the same electrolytes with 10 and
50 vol\% PAAM. Solid lines: our calculation results within the
three-shell approximation, given by Eq.~(\ref{eq: effconduct}) at
$M=3$, under the suggestion that the conductivities of the
constituents obey the VTF equation (\ref{eq:VTF}) with parameter
values summarized in Table~\ref{tab:temp}. Dotted lines, (b): the
same, but for multiplicatively renormalized $\sigma_{\rm eff}$:
$0.40\, \sigma_{\rm eff}$ and $0.75\, \sigma_{\rm eff}$ for
electrolytes with 10 and 50  vol\% PAAM, respectively. The
percentage deviations of the above data from and $R^2$ values for
the calculated (and, if so, renormalized) curves are presented in
Fig.~\ref{fig:RAE-isotherm}.}
\end{figure}

\begin{figure}[htb]
\centering
\includegraphics[width=0.85\textwidth]{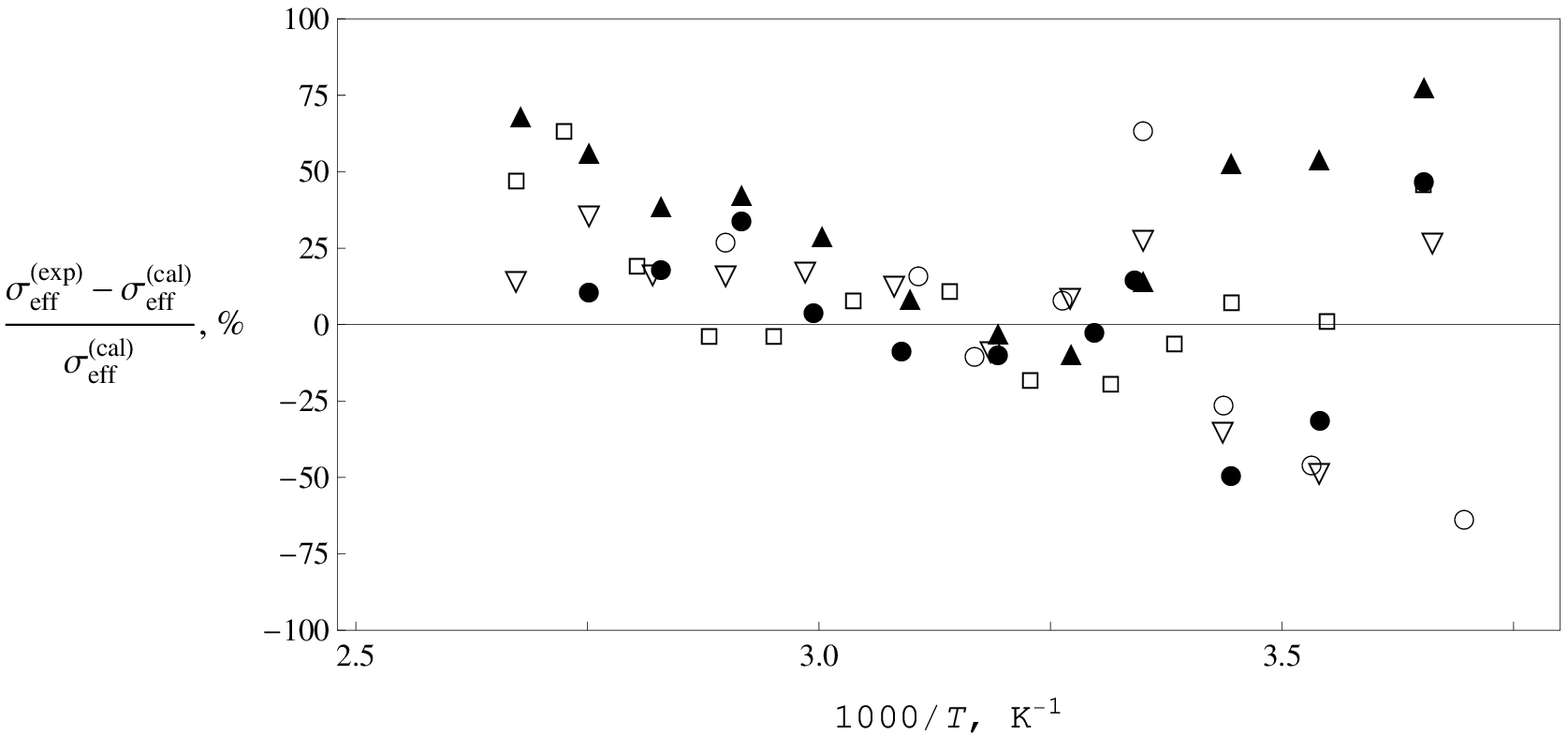}
\caption{\label{fig:RAE-isotherm}  Percentage deviations of
experimental $\sigma_{\rm eff}$ vs $T$ data~\cite{Wiec1994} for
OMPEO--PAAM--LiClO$_4$ electrolytes (10 mol\% LiClO$_4$, after
annealing) from the calculated (and, if so, renormalized) curves
in Fig.~\ref{fig:OMPEO-LiClO4-TempDependence}. The deviations are
calculated for all the electrolyte samples with 5 ($\nabla$), 10
($\circ$), 25 ($\square$), 40 ($\blacktriangle$), and 50
($\bullet$) vol\% PAAM. The $R^2$ values for the indicated curves
are 94.8, 94.0, 83.4, 77.5, and 96.0\,\%, respectively.}
\end{figure}

The results of applying the three-shell approximation to three
$\sigma_{\rm{eff}}$ vs $c$ isotherms \cite{Wiec1994} of
OMPEO--PAAM--LiClO$_4$ electrolytes  (with 10 mol\% LiClO$_4$,
after annealing) are presented in
Figs.~\ref{fig:OMPEO-LiClO4-Temp},
\ref{fig:OMPEO-LiClO4-Temp-Deviation}  and
Table~\ref{tab:isotherms}; the temperature-independent values
$\delta_1=0.40$, $\delta_2$=0.80, and $\delta_3=1.40$ were used
(see Table~\ref{tab:adjustable_params}). The fitting values
obtained for $\sigma_0$ and $\sigma_{2,j}$ were then used to
estimate the VTF parameters for the CPE's constituents; they are
summarized in Table~\ref{tab:temp}. Finally, employing
Eqs.~(\ref{eq: effconduct}) at $M=3$ and (\ref{eq:VTF}) and the
above geometrical and electrical data for the CPE's constituents,
the $\sigma_{\rm{eff}}$ vs $T$  dependences were recovered for all
OMPEO--PAAM--LiClO$_4$ samples discussed in \cite{Wiec1994}; these
results are shown in Figs.~\ref{fig:OMPEO-LiClO4-TempDependence}
and \ref{fig:RAE-isotherm}.

Three remarks are worth making here:

1. Our estimates $B= 1270\,\rm{K}$ and $ T_0 =190\, \rm{K}$ for
pure OMPEO turn out to be close to estimates  $B= 1200\,\rm{K}$
and $ T_0 =195\, \rm{K}$ obtained from direct conductivity
measurements \cite{Wiec1994}. In contrast, our estimate for the
preexponential factor $A$ differs noticeably from that in
\cite{Wiec1994}: $A =36.1$ and $ 27.0\, \rm{S}\cdot
\rm{K}^{1/2}/\rm{cm}$, respectively. Together with the fact that
our theoretical curves fit the experimental data better, this
result may indicate that the effective electrical properties of
the polymer matrix are altered in the course of CPE preparation.

2. All our estimates of the VTF parameters for the shells fall in
the value ranges reported in \cite{Wiec1994} for all
OMPEO--PAAM--LiClO$_4$ samples studied. Therefore, from this point
of view, they are not contradictory.

3. Taking into account the original uncertainties in the shells'
conductivity values obtained by processing the three conductivity
isotherms, it can be concluded that the experimental data for the
samples with 5, 25, and 40 vol\% PAAM are reproduced by our theory
well. Those for the samples with 10 and 50 vol\% PAAM are
recovered sufficiently well; the agreement is improved by a
multiplicative renormalization (multiplication by a constant
factor) of the theoretical results. The latter fact may be
attributed to the discrepancy in the $A$-values obtained in
\cite{Wiec1994} and in the present work for the matrix
conductivity.

\begin{table*}[htb]
\centering
\begin{threeparttable}
\caption{\label{tab:isotherms} Conductivity values, in S/cm, used
to fit $\sigma_{\rm{eff}}$ vs $c$ isotherms \cite{Wiec1994} for
OMPEO--PAAM--LiClO$_4$ electrolytes \tnote{a,b}  (see
Fig.~\ref{fig:OMPEO-LiClO4-Temp}). }

\begin{tabular}{llll}
\hline
\textrm{Constituent}   & $t = 0\,\rm { ^oC}$ & $t = 25\, \rm {^oC}$ & $t = 100\, \rm {^oC}$ \\
\hline
Matrix, $\sigma_0$           &  $4.64\times 10^{-7}$  &  $1.57\times 10^{-5}$  &  $1.78\times 10^{-3}$   \\
First shell, $\sigma_{21}$   &  $5.75\times 10^{-7}$  &  $8.70\times 10^{-6}$  &  $4.21\times 10^{-4}$    \\
Second shell, $\sigma_{22}$  &  $1.025\times 10^{-3}$ &  $7.74\times 10^{-3}$  &  $1.00\times 10^{-1}$   \\
Third shell, $\sigma_{23}$   &  $1.07\times 10^{-7}$  &  $3.12\times 10^{-6}$  &  $1.36\times 10^{-4}$ \\
\hline
\end{tabular}
\begin{tablenotes}
\item[a] With 10 mol\% LiClO$_4$. \item[b]  Due to the
complexation of $\rm{Li}^+ $ cations with PAAM chains,
PAAM--LiClO$_4$ cores are essentially nonconductive, with room
temperature conductivity $\sigma_1\sim 1\times 10^{-12}\, \rm
{S/cm}$ \cite{Wiec1994}. This  value was used in our calculations.
An increase of $\sigma_1$ by several orders of magnitude does not
affect, within the required accuracy, the results obtained.
\end{tablenotes}
\end{threeparttable}
\end{table*}

\begin{table*}[htb]
\centering
\begin{threeparttable} \caption{\label{tab:temp} VTF parameters obtained for
OMPEO--PAAM--LiClO$_4$ electrolytes \tnote{a}}

\begin{tabular}{llll}
\hline
\textrm{Constituent}   & $A$, $\rm{S}\times \rm{K}^{1/2}/\rm{cm}$ & $B$, K & $T_0$, K  \\
\hline
Matrix, $\sigma_0$           &  36.1\tnote{a}  &  1270  &  190   \\
First shell, $\sigma_{21}$   &  4.33  &  1210   &  180    \\
Second shell, $\sigma_{22}$  &  71.1   &  634     &  197   \\
Third shell, $\sigma_{23}$   & 0.229   &  720   &  212 \\
\hline
\end{tabular}
\begin{tablenotes}
\item[a] With 10 mol\% LiClO$_4$.
\end{tablenotes}
\end{threeparttable}
\end{table*}

\section{\label{sec:concnlusion}Conclusion}

We have proposed a new  approach to finding the effective
quasistatic electrical conductivity $\sigma_{\rm eff}$ of CPEs.
Its two major features are as follows:

1. The microstructure of a CPE is viewed as a result of embedding
hard-core particles with adjacent penetrable (freely overlapping)
layers into a uniform matrix. The layers are isotropic and, in
general, inhomogeneous: they comprise a finite or infinite number
of concentric shells. With the order of dominance taken to be hard
cores $>$ shorter-radius shells $>$ longer-radius shell $>$ the
matrix, the local value of the complex permittivity and,
consequently, that of the quasistatic electrical conductivity in
the system are determined by the distance from the point of
interest to the center of the nearest particle. The layer's
conductivity profile is expected to account for different
mechanisms contributing to the electrical properties of real CPEs.
Some of its parts should have enhanced ionic conductivity, as
compared to that of the polymer-based matrix with a higher degree
of crystallinity, to form percolation paths.

2. The effective complex permittivity and then $\sigma_{\rm eff}$
are found  for the model by applying the CGA. This approach makes
it possible to avoid a detailed modeling of the many-particle
polarization and correlation process in the system. The desired
$\sigma_{\rm eff}$ is eventually shown to satisfy an integral
relation found by summing up all the moments of the local complex
permittivity deviations from the effective complex permittivity
and passing to the quasistatic limit.

Putting the model to the test reveals that it is capable of
recovering extensive experimental data \cite{Przl1995,Wiec1994}
for a series of PEO- and OMPEO-based electrolytes in a wide range
of concentration. However, the agreement between the theory and
experiment is reachable only under the suggestion of electrical
inhomogeneity of the layers. To reproduce the major features of
the concentration behavior of $\sigma_{\rm eff}$ for those data,
the two-shell structures for $\sigma_2(r)$ are sufficient to be
used for PEO--NaI--NASICON ~\cite{Przl1995} and
(PEO)$_{10}$--NaI--$\theta$-Al$_2$O$_3$~\cite{Wiec1994}), and the
three-shell ones for PEO--PAAM--LiClO$_4$~\cite{Przl1995,
Wiec1994} and OMPEO--PAAM--LiClO$_4$~\cite{Wiec1994}. The
continuous sigmoid-type analogues of these conductivity profiles
are also  proposed. Their shapes are similar to those suggested in
simulations~\cite{Knauth2008,Siekierski2005,Siekierski2006}.
Finally,  finding the VTF parameters for the polymer matrix and
shell conductivities obtained by processing the three available
conductivity isotherms within the three-shell approximation, the
temperature behavior of $\sigma_{\rm  eff}$ for different samples
of OMPEO--PAAM--LiClO$_4$~\cite{Wiec1994} is recovered
satisfactorily.

The fitting results indicate that for the above CPEs, the observed
behavior of $\sigma_{\rm eff}$ can be attributed to several
mechanisms: (1) a change of the matrix's conductivity in the
course of preparation of a CPE; (2) an amorphization of the
polymer matrix by filler grains; (3) a stiffening effect of the
filler grains on the amorphous phase; (4) effects caused by
irregularities in the filler grains' shapes; (5) a formation of a
highly-resistive polymer-filler interface. These mechanisms are
effectively taken into account through {the parameters of} the
outermost [mechanism (1)], central [mechanism (2)], and innermost
[mechanisms (3) and (4)] parts in $\sigma_2(r)$, which gradually
come into play as $c$ is increased; or a reduced value, compared
to that in a solitary state,  of the conductivity for
highly-conductive filler grains inside CPEs [mechanism (5)].

\section*{Acknowledgment}
We are grateful to an anonymous Referee for constructive and
stimulating remarks.

\newpage


\begin{thebibliography}{10}
\expandafter\ifx\csname url\endcsname\relax
  \def\url#1{\texttt{#1}}\fi
\expandafter\ifx\csname
urlprefix\endcsname\relax\def\urlprefix{URL }\fi
\expandafter\ifx\csname href\endcsname\relax
  \def\href#1#2{#2} \def\path#1{#1}\fi

\bibitem{Ploch1988}
J.~P\l{}ocharski, W.~Wieczorek, {PEO} based composite solid
electrolyte
  containing {NASICON}, Solid State Ionics 28--30 (1988) 979--982.
\newblock \href {http://dx.doi.org/10.1016/0167-2738(88)90315-3}
  {\path{doi:10.1016/0167-2738(88)90315-3}}.

\bibitem{Przl1995}
J.~Przyluski, M.~Siekierski, W.~Wieczorek, Effective medium theory
in studies
  of conductivity of composite polymeric electrolytes, Electrochimica Acta 40
  (1995) 2101--2108.
\newblock \href {http://dx.doi.org/10.1016/0013-4686(95)00147-7}
  {\path{doi:10.1016/0013-4686(95)00147-7}}.

\bibitem{Wiec1989}
W.~Wieczorek, K.~Such, H.~Wyci\'slik, J.~P\l{}ocharski,
Modifications of
  crystalline structure of {PEO} polymer electrolytes with ceramic additives,
  Solid State Ionics 36 (1989) 255--257.
\newblock \href {http://dx.doi.org/10.1016/0167-2738(89)90185-9}
  {\path{doi:10.1016/0167-2738(89)90185-9}}.

\bibitem{Wiec1994}
W.~Wieczorek, K.~Such, Z.~Florjanczyk, J.~Stevens, Polyether,
polyacrylamide,
  $\rm {LiClO_4}$ composite electrolytes with enhanced conductivity, J. Phys.
  Chem. 98 (1994) 6840--6850.
\newblock \href {http://dx.doi.org/10.1021/j100078a029}
  {\path{doi:10.1021/j100078a029}}.

\bibitem{Moh1998}
M.~A. Moharram, M.~A. Soliman, H.~M. El-Gendy, Electrical
conductivity of
  poly(acrylic acid) -- polyacrylamide complexes, J. Appl. Polymer Sci. 68
  (1998) 2049--2055.
\newblock \href
  {http://dx.doi.org/10.1002/(SICI)1097-4628(19980620)68:12<2049::AID-APP19>3.0.CO;2-W}
  {\path{doi:10.1002/(SICI)1097-4628(19980620)68:12<2049::AID-APP19>3.0.CO;2-W}}.

\bibitem{Zal1996}
A.~Zalewska, W.~Wieczorek, J.~R. Stevens, Composite polymeric
electrolytes from
  the $\rm {PEO-PAAM-NH_4SCN}$ system, J. Phys. Chem. 100 (1996) 11382--11388.
\newblock \href {http://dx.doi.org/10.1021/jp952909n}
  {\path{doi:10.1021/jp952909n}}.

\bibitem{Siekierski2007}
M.~Siekierski, K.~Nadara, Mesoscale models of {AC} conductivity in
composite
  polymeric electrolytes, J. Pow. Sour. 173 (2007) 748--754.
\newblock \href {http://dx.doi.org/10.1016/j.jpowsour.2007.05.063}
  {\path{doi:10.1016/j.jpowsour.2007.05.063}}.

\bibitem{MacCallum19871989}
J.~R. MacCallum, C.~Vincent (Eds.), Polymer Electrolyte Reviews,
vol. 1, 1987
  and vol. 2, 1989, Elsevier Applied Science, London.

\bibitem{Scrosati1993}
B.~Scrosati (Ed.), Applications of Electroactive Polymers, Chapman
\& Hall,
  London, 1993.

\bibitem{Bruce1995}
P.~Bruce (Ed.), Solid State Electrochemistry, Cambridge University
Press,
  Cambridge, 1995.

\bibitem{Quartarone1998}
E.~Quartarone, P.~Mustarelli, A.~Magistris, {PEO}-based composite
polymer
  electrolytes, Solid State Ionics 110 (1998) 1--14.
\newblock \href {http://dx.doi.org/10.1016/S0167-2738(98)00114-3}
  {\path{doi:10.1016/S0167-2738(98)00114-3}}.

\bibitem{Song1999}
J.~Y. Song, Y.~Y. Wang, C.~Wan, Review of gel-type polymer
electrolytes for
  lithium-ion batteries, J. Pow. Sour. 77 (1999) 183--197.
\newblock \href {http://dx.doi.org/10.1016/S0378-7753(98)00193-1}
  {\path{doi:10.1016/S0378-7753(98)00193-1}}.

\bibitem{Croce2000}
F.~Croce, L.~Persi, F.~Ronci, B.~Scrosati, Nanocomposite polymer
electrolytes
  and their impact on the lithium battery technology, Solid State Ionics 135
  (2000) 47--52.
\newblock \href {http://dx.doi.org/10.1016/S0167-2738(00)00329-5}
  {\path{doi:10.1016/S0167-2738(00)00329-5}}.

\bibitem{Sequeira2010}
C.~Sequeira, D.~Santos (Eds.), Polymer Electrolytes. Fundamentals
and
  Applications, Woodhead Publishing, Oxford, 2010.

\bibitem{Tarascon2015}
J.-M. Tarascon, P.~Simon, Electrochemical Energy Storage, Vol.~1,
ISTE and John
  Wiley \& Sons, 2015.

\bibitem{Knauth2008}
W.~Wieczorek, M.~Siekierski, Nanocomposites. Ionic Conducting
Materials and
  Structural Spectroscopies, Springer, New York, 2008, Ch. Composite Polymeric
  Electrolytes, pp. 1--70.

\bibitem{PrzlSymp1988}
J.~Przy\l{}uski, et~al., Proton conducting polymeric electrolytes
from poly
  (ethyleneoxide) system, in: B.~V.~R. Chowdari, Q.~Liu, L.~Chen (Eds.), Recent
  Advances in Fast Ion Conducting Materials and Devices. Proc. 2nd Asian
  Conference on Solid State lonics, World Scientific, Singapore, 1990, p. 307.

\bibitem{Bulavin2015}
L.~A. Bulavin, I.~A. Melnyk, A.~I. Goncharuk, V.~V. Klepko, N.~I.
Lebovka,
  E.~A. Lysenkov, Effect of molecular weight on the properties of polyethylene
  glycol doped by multiwalled carbon nanotubes, Dopov. Nac. akad. nauk Ukr. 8
  (2015) 72--78.

\bibitem{croce1992}
F.~Croce, B.~Scrosati, M.~G., Electrochemical and spectroscopic
study of the
  transport properties of composite polymer electrolytes, Chem. Mater. 4 (1992)
  1134--1136.
\newblock \href {http://dx.doi.org/10.1021/cm00024a003}
  {\path{doi:10.1021/cm00024a003}}.

\bibitem{Ploch1989}
J.~P\l{}ocharski, W.~Wieczorek, J.~Przy\l{}uski, K.~Such, Mixed
solid
  electrolytes based on poly(ethylene oxide), Appl. Phys. A 49 (1989) 55--60.
\newblock \href {http://dx.doi.org/10.1007/BF00615464}
  {\path{doi:10.1007/BF00615464}}.

\bibitem{Przyluski1990}
J.~Przy\l{}uski, K.~Such, H.~Wyci\'{s}lik, W.~Wieczorek,
Z.~Floria\'{n}czyk,
  {PEO}--based polymer blends as materials for solid electrolytes, Synth. Met.
  35 (1989) 241--247.
\newblock \href {http://dx.doi.org/10.1016/0379-6779(90)90048-P}
  {\path{doi:10.1016/0379-6779(90)90048-P}}.

\bibitem{Nan1993}
C.-W. Nan, Physics of inhomogeneous inorganic materials, Prog.
Mater. Sci. 37
  (1993) 1--116.
\newblock \href {http://dx.doi.org/10.1016/0079-6425(93)90004-5}
  {\path{doi:10.1016/0079-6425(93)90004-5}}.

\bibitem{Jiang1995a}
S.~Jiang, J.~B. Wagner, A theoretical model for composite
electrolytes -- {I.}
  {Space} charge layer as a cause for charge--carrier enhancement, J. Phys.
  Chem. Solids 56 (1995) 1101--1111.
\newblock \href {http://dx.doi.org/10.1016/0022-3697(95)00025-9}
  {\path{doi:10.1016/0022-3697(95)00025-9}}.

\bibitem{Jiang1995b}
S.~Jiang, J.~B. Wagner, A theoretical model for composite
electrolytes -- {II}.
  {P}ercolation model for ionic conductivity enhancement, J. Phys. Chem.
  Solids. 56 (1995) 1113--1124.
\newblock \href {http://dx.doi.org/10.1016/0022-3697(95)00026-7}
  {\path{doi:10.1016/0022-3697(95)00026-7}}.

\bibitem{Tod03}
M.~G. Todd, F.~G. Shi, Characterizing the interphase dielectric
constant of
  polymer composite materials: Effect of chemical coupling agents, J. Appl.
  Phys. 94 (2003) 4551--4557.
\newblock \href {http://dx.doi.org/10.1063/1.1604961}
  {\path{doi:10.1063/1.1604961}}.

\bibitem{Maxwell1873}
J.~C. Maxwell, A Treatise on Electricity and Magnetism, 1st
Edition, Vol.~1,
  Clarendon Press, Oxford, 1873, pp. 362--365.

\bibitem{Maxwell1904}
J.~C.~M. Garnett, Colours in metal glasses and metalic films,
Trans. R. Soc.
  Lond. A 203 (1904) 385.
\newblock \href {http://dx.doi.org/10.1098/rsta.1904.0024}
  {\path{doi:10.1098/rsta.1904.0024}}.

\bibitem{Landau1982}
L.~D. Landau, E.~M. Lifshitz, L.~P. Pitaevskii, Course of
Theoretical Physics.
  Electrodynamics of Continuous Media, Vol.~8, Pergamon, Oxford, 1984.

\bibitem{Bruggeman1935a}
D.~Bruggeman, Berechnnung verschiedener physikalischer konstanten
von
  heterogenen substanzen 1. dielektrizit\"{a}tskonstanten und
  leitf\"{a}higkeiten der mischk\"{o}rper aus isotropen substanzen, Ann. Phys.
  (Leipzig) 24 (1935) 636--664.
\newblock \href {http://dx.doi.org/10.1002/andp.19354160705}
  {\path{doi:10.1002/andp.19354160705}}.

\bibitem{Landauer1952}
R.~Landauer, The electrical resistance of binary metallic
mixtures, J. Appl.
  Phys. 23 (1952) 779.
\newblock \href {http://dx.doi.org/10.1063/1.1702301}
  {\path{doi:10.1063/1.1702301}}.

\bibitem{Nakamura1982}
M.~Nakamura, A method to improve the effective medium theory
towards
  percolation problem, J. Phys. C: Solid State Phys. 15 (1982) L749--752.
\newblock \href {http://dx.doi.org/10.1088/0022-3719/15/23/005}
  {\path{doi:10.1088/0022-3719/15/23/005}}.

\bibitem{Nakamura1984}
M.~Nakamura, Conductivity for the site-percolation problem by an
improved
  effective-medium theory, Phys. Rev. B 29 (1984) 3691--3693.
\newblock \href {http://dx.doi.org/10.1103/PhysRevB.29.3691}
  {\path{doi:10.1103/PhysRevB.29.3691}}.

\bibitem{Nan1991L}
C.-W. Nan, D.~M. Smith, On comment on ``{E}nhancement of ionic
conduction in
  {CaF}$_2$ and {BaF}$_2$ by dispersion of {Al}$_2${O}$_3$'', J. Mat. Sci.
  Lett. 10 (1991) 1142--1143.
\newblock \href {http://dx.doi.org/10.1007/BF00744107}
  {\path{doi:10.1007/BF00744107}}.

\bibitem{Nan1991}
C.-W. Nan, D.~M. Smith, A.c. electrical properties of composite
solid
  electrolytes, Mat. Sci. Eng. B 10 (1991) 99--106.
\newblock \href {http://dx.doi.org/10.1016/0921-5107(91)90115-C}
  {\path{doi:10.1016/0921-5107(91)90115-C}}.

\bibitem{Siekierski2005}
M.~Siekierski, K.~Nadara, Modeling of conductivity in composites
with random
  resistor networks, Electrochimica Acta 50 (2005) 3796--3804.
\newblock \href {http://dx.doi.org/10.1016/j.electacta.2005.02.046}
  {\path{doi:10.1016/j.electacta.2005.02.046}}.

\bibitem{Siekierski2006}
M.~Siekierski, K.~Nadara, P.~Rzeszotarski, Conductivity simulation
in composite
  polymeric electrolytes, J. New Mat. Electrochem. Systems 9 (2006) 375--390.
\newblock \href {http://dx.doi.org/10.1.1.455.1852}
  {\path{doi:10.1.1.455.1852}}.

\bibitem{Sushko2007}
M.~Y. Sushko, Dielectric permittivity of suspensions, Zh. Eksp.
Teor. Fiz. 132
  (2007) 478--484, [JETP \textbf{105} (2007) 426--431].
\newblock \href {http://dx.doi.org/10.1134/S106377610}
  {\path{doi:10.1134/S106377610}}.

\bibitem{Sushko2009}
M.~Y. Sushko, S.~K. Kris'kiv, Compact group method in the theory
of
  permittivity of heterogeneous systems, Zh. Tekh. Fiz. 79 (2009) 97--101,
  [Tech. Phys. \textbf{54} (2009) 423--427].
\newblock \href {http://dx.doi.org/10.1134/S1063784209030165}
  {\path{doi:10.1134/S1063784209030165}}.

\bibitem{Sushko22009}
M.~Y. Sushko, Effective permittivity of mixtures of anisotropic
particles, J.
  Phys. D: Appl. Phys. 42 (2009) 155410.
\newblock \href {http://dx.doi.org/10.1088/0022-3727/42/15/155410}
  {\path{doi:10.1088/0022-3727/42/15/155410}}.

\bibitem{Sushko2017}
M.~Y. Sushko, Effective dielectric response of dispersions of
graded particles,
  Phys. Rev. E 96 (2017) 062121.
\newblock \href {http://dx.doi.org/10.1103/PhysRevE.96.062121}
  {\path{doi:10.1103/PhysRevE.96.062121}}.

\bibitem{Sushko2013}
M.~Y. Sushko, A.~K. Semenov, Conductivity and permittivity of
dispersed systems
  with penetrable particle-host interphase, Cond. Matter Phys. 16 (2013) 13401.
\newblock \href {http://dx.doi.org/10.5488/CMP.16.13401}
  {\path{doi:10.5488/CMP.16.13401}}.

\bibitem{Sushko2016}
M.~Y. Sushko, V.~Y. Gotsulskiy, M.~V. Stiranets, Finding the
effective
  structure parameters for suspensions of nano-sized insulating particles from
  low-frequency impedance measurements, J. Mol. Liq. 222 (2016) 1051--1060.
\newblock \href {http://dx.doi.org/10.1016/j.molliq.2016.07.021}
  {\path{doi:10.1016/j.molliq.2016.07.021}}.

\bibitem{Tomylko2015}
S.~Tomylko, O.~Yaroshchuk, N.~Lebovka, Two-step electrical
percolation in
  nematic liquid crystals filled with multiwalled carbon nanotubes, Phys. Rev.
  E 92 (2015) 012502.
\newblock \href {http://dx.doi.org/10.1103/PhysRevE.92.012502}
  {\path{doi:10.1103/PhysRevE.92.012502}}.

\bibitem{Semenov2018}
A.~K. Semenov, On applicability of differential mixing rules for
statistically
  homogeneous and isotropic dispersions, J. Phys. Commun. 2 (2018) 035045.
\newblock \href {http://dx.doi.org/10.1088/2399-6528/aab060}
  {\path{doi:10.1088/2399-6528/aab060}}.

\bibitem{Rikvold1985}
P.~A. Rikvold, G.~Stell, Porosity and specific surface for
  interpenetrablesphere models of twophase random media, J. Chem. Phys. 82
  (1985) 1014--1020.
\newblock \href {http://dx.doi.org/10.1063/1.448966}
  {\path{doi:10.1063/1.448966}}.

\bibitem{Rikvold1985a}
P.~A. Rikvold, G.~Stell, {\it D}-dimensional
interpenetrable-sphere models of
  random two-phase media: Microstructure and an application to chromatography,
  J. Colloid Interface Sci. 108 (1985) 158--173.
\newblock \href {http://dx.doi.org/10.1016/0021-9797(85)90246-2}
  {\path{doi:10.1016/0021-9797(85)90246-2}}.

\bibitem{Lee88}
S.~B. Lee, S.~Torquato, Porosity for the
penetrable-concentric-shell model of
  two-phase disordered media: Computer simulation results, J. Chem. Phys. 89
  (1988) 3258--3263.
\newblock \href {http://dx.doi.org/10.1063/1.454930}
  {\path{doi:10.1063/1.454930}}.

\bibitem{Rottereau03}
M.~Rottereau, J.~C. Gimel, T.~Nicolai, D.~Durand, 3d {M}onte
{C}arlo simulation
  of site-bond continuum percolation of spheres, Eur. Phys. J. E 11 (2003)
  61--64.
\newblock \href {http://dx.doi.org/10.1140/epje/i2003-10006-x}
  {\path{doi:10.1140/epje/i2003-10006-x}}.

\end{thebibliography}
\end{document}